\documentclass{aa}  

\usepackage{graphicx}
\usepackage{txfonts}
\usepackage{xcolor}
\usepackage{keyval}
\usepackage{hyperref}
\begin{document}

   \title{Luminous Fast Blue Optical Transients as very massive star core-collapse events}

   \author{A. A. Chrimes
          \inst{1,2}\fnmsep\thanks{ESA Research Fellow}
          \and
          P. G. Jonker\inst{2}
          \and
          A. J. Levan\inst{2,3}
          \and
          A. Mummery\inst{4}
          }

   \institute{European Space Agency (ESA), European Space Research and Technology Centre (ESTEC), Keplerlaan 1, 2201 AZ Noordwijk, the Netherlands \\
              \email{ashley.chrimes@esa.int}
         \and
            Department of Astrophysics/IMAPP, Radboud University, PO Box 9010, 6500 GL Nijmegen, The Netherlands
         \and
            Department of Physics, University of Warwick, Gibbet Hill Road, CV4 7AL Coventry, United Kingdom 
        \and 
            School of Natural Sciences, Institute for Advanced Study, 1 Einstein Drive, Princeton, NJ 08540, USA 
            }

   \date{Received October 3, 2025; accepted XXX}

 
  \abstract
   {Luminous Fast Blue Optical Transients (LFBOTs) are rare extragalactic events of unknown origin. Tidal disruptions of white dwarfs by intermediate mass black holes, mergers of black holes and Wolf-Rayet stars, and failed supernovae are among the suggested explanations.}
   {In this paper, we explore the viability of very massive star core-collapse events as the origin of LFBOTs. The appeal of such a model is that the formation of massive black holes via core collapse may yield observational signatures that can match the disparate lines of evidence that point towards both core-collapse and tidal disruption origins for LFBOTs.}
   {We explore the formation rate of massive black holes in binary population synthesis models, and compare the metallicities of their progenitors with the observed metallicities of LFBOT host galaxies. We further examine the composition, mass loss rates and fallback masses of these stars, placing them in the context of LFBOT observations.}
   {The formation rate of black holes with masses greater than $\sim$30--40\,M$_{\odot}$ is found to be similar to the observed LFBOT rate. The stars producing these black holes are biased to low metallicity (Z$<$0.3\,Z$_{\odot}$), are H and He-poor and have dense circumstellar media. However, some LFBOTs have host galaxies with higher metallicities than predicted, and they typically have denser local environments (plausibly due to late stage mass loss not captured in the models). We find that long-lived emission from an accretion disc (as implicated in the prototypical LFBOT AT\,2018cow) can only be produce in these events under maximal disc mass and angular momentum conditions.}
   {We conclude that (very) massive star core-collapse is a plausible explanation for at least some LFBOTs, but faces challenges. The preferred progenitors for LFBOTs in the failed supernova interpretation overlap with those predicted to produce super-kilonovae. We therefore suggest that LFBOTs are promising targets to search for super-kilonovae, and that they may contribute non-negligibly to the r-process enrichment of galaxies.}


   \titlerunning{LFBOTs as very massive star core-collapse events}
   \authorrunning{A. A. Chrimes et al.}
   \maketitle
%

\section{Introduction}
Luminous Fast Blue Optical Transients (LFBOTs), or Cow-like transients after the prototypical AT\,2018cow \citep[][]{2018ApJ...865L...3P,2019MNRAS.484.1031P}, are rapidly evolving transients with peak optical-ultraviolet luminosities rivalling superluminous supernovae (SLSNe). The early optical-UV spectra are well described by a hot, featureless blackbody (T$>$20000\,K), lacking H and He lines initially. They have decay rates of $\gtrsim$0.3 magnitudes/day, ruling out a significant contribution from $^{56}$Ni-powered emission. They are also accompanied by bright X-ray (and at late times, radio) emission. The late-time radio emission is attributed to self-absorbed synchrotron emission from an expanding blastwave in a dense circumstellar medium (CSM). 

A modest sample of LFBOTs has now been identified in addition to AT\,2018cow: ZTF18abvkwla \citep[AT2018lug,][]{2020ApJ...895...49H}, CSS161010 \citep{2020ApJ...895L..23C,2024ApJ...977..162G}, ZTF20acigmel \citep[AT2020xnd,][]{2021MNRAS.508.5138P,2022ApJ...926..112B}, AT2020mrf \citep{2022ApJ...934..104Y}, AT2022tsd \citep{2023RNAAS...7..126M}, AT2023fhn \citep{2024MNRAS.527L..47C,2024AandA...691A.329C} and AT2024wpp \citep{2025MNRAS.537.3298P,2025ApJ...993L...6N,2025arXiv250900952N}, as well as the slower-evolving and radio-faint, but otherwise similar, AT2024puz \citep{2025ApJ...995..228S} and several other candidates which lack extensive multi-wavelength follow-up. Some transients display very blue colours, evolve rapidly at early times and produce bright X-ray emission, similar to LFBOTs, but are followed by broad lined type Ic SNe \citep[e.g.\ the Einstein Probe-discovered transients EP2404014A and EP250108A,][]{2025ApJ...982L..47V,2025ApJ...988L..14E,2025ApJ...988L..13R}. However, these events likely do not share the same progenitors as LFBOTs, given that they successfully launched supernovae, and we henceforth focus solely on the SN-less, Cow-like LFBOTs, whose origins remain unknown. The volumetric rate of LFBOTs is estimated to be $\sim$0.1\% of the core-collapse SN rate \citep{2023ApJ...949..120H}. 

Two categories of model to explain LFBOTs have emerged as most likely \citep[e.g.][]{2019MNRAS.484.1031P}. The first invokes tidal disruption events (TDEs) of H and He poor, compact stars (e.g. a white dwarfs) by intermediate mass black holes \citep[IMBHs,][]{2019MNRAS.487.2505K}. This model is challenged primarily due to the dense CSM observed in LFBOTs. Although such a environment is possible in TDEs involving stars disrupted by supermassive black holes, it is less clear whether white dwarf plus IMBH TDEs could produce similar circumstellar density profiles \citep{2019ApJ...872...18M,2024ApJ...974...67L}. The second category of proposals are broadly themed around stellar mass black hole accretion, either in mergers/collisions \citep{2022ApJ...932...84M,2025Ap&SS.370...11G,2025ApJ...986...84T,2025arXiv251009745K} or core-collapse events \citep[e.g.][]{2015MNRAS.451.2656K,2019ApJ...872...18M,2019MNRAS.485L..83Q,2022MNRAS.513.3810G}.

At least some sufficiently massive stars are expected to directly form black holes without launching a supernova. The landscape of explodability is thought to be complex, with neutron star and some black hole-forming events capable of launching successful supernovae, and other black hole-forming core-collapses resulting in direct black hole formation without any significant ejecta \citep[e.g.][]{2003ApJ...591..288H,2020ApJ...890...51E}. Observational evidence for `quiet' black hole formation comes from a lack of massive stars $\gtrsim 20$\,M$_{\odot}$ in pre-supernovae imaging, and the disappearance of supergiant stars without evidence for a supernova (\citealt{2009MNRAS.395.1409S,2014ApJ...791..105W,2015MNRAS.453.2885R,2017MNRAS.468.4968A,2017MNRAS.469.1445A,2024ApJ...964..171B}). 

Some LFBOT models invoke such a scenario, where a massive star collapses to a black hole without a successful supernova (a `failed supernova'), followed by the accretion of some fraction of the remaining (`leftover') material onto the nascent BH \citep[e.g.][]{2014ApJ...781..119P,2015MNRAS.451.2656K,2018MNRAS.476.2366F,2019MNRAS.485L..83Q}. The failed supernova hypothesis resembles early collapsar gamma-ray burst (GRB) models \citep[e.g.][]{1993ApJ...405..273W,1999ApJ...524..262M,1999ApJ...526..152F}, which also feature accreting black holes born in core-collapse, but without a successful (or a weak) supernova. In LFBOTs there is no evidence for a supernova, but unlike in collapsar GRBs, no gamma-ray emission has been observed, and the outflows are only mildly relativistic \citep[e.g.][]{2020ApJ...895L..23C}.

While some fast evolving optical transients can be explained solely with circumstellar interactions \citep{2022ApJ...926..125P}, in LFBOTs there is evidence for central engine activity. AT\,2018cow showed short timescale variability in its light-curve \citep{2019ApJ...871...73H,2019ApJ...872...18M,2022NatAs...6..249P}, as well as UV and X-ray emission lasting at least several years post-explosion \citep{2022MNRAS.512L..66S,2023MNRAS.519.3785S,2023MNRAS.525.4042I,2023ApJ...955...43C,2024ApJ...963L..24M,2025MNRAS.544L.108I}. Magnetar central engines have been proposed \citep{2019ApJ...878...34F,2020ApJ...888L..24M,2022ApJ...935L..34L,2024ApJ...963L..13L}, but struggle to explain both the early-time emission and late-time optical/UV plateau \citep{2023ApJ...955...43C}. Assuming instead that LFBOTs are black hole (BH) powered, estimates for the BH mass in such a scenario have been derived from quasi period oscillations \citep{2022NatAs...6..249P} and accretion disc modelling \citep{2023MNRAS.525.4042I,2024ApJ...963L..24M,2024A&A...691A.228C,2025MNRAS.544L.108I} yielding masses which span the upper end of the stellar mass BH distribution into the intermediate BH mass regime. The observation of late-time, short duration giant optical flares from AT\,2022tsd \citep{2023Natur.623..927H} may also be consistent with a highly variable accretion rate, either around the black hole formed in the event or a companion black hole \citep{2024ApJ...972L..17L}. Similar flaring behaviour has not been found in AT20224wpp \citep{2025ApJ...993...76O}.

Black hole mass estimates in LFBOTs are high, and they have thus far been discovered in non-nuclear regions of low-metallicity, star-forming galaxies. The aim of this paper is to explore other observational consequences - for example in terms of volumetric rates and the metallicities of their hosts - if LFBOTs are indeed powered by massive BH formation in failed supernovae. In Section \ref{sec:rates} we compare population synthesis predictions for the rate of BH formation as a function of mass with the observed LFBOT rate. We then investigate in Section \ref{sec:hosts} whether the metallicity bias associated with the formation of the most massive stellar mass BHs is consistent with LFBOT environments. We explore in Section \ref{sec:emiss} whether the selected stellar models can plausibly reproduce the key characteristics of LFBOT emission, with a discussion and conclusions following in Sections \ref{sec:disc} and \ref{sec:conc}.

\section{Black hole formation and LFBOT event rates}\label{sec:rates}
We first ask the question: what is the volumetric formation rate of stellar-mass, core-collapse black holes with mass $>$\,M$_{\rm BH, min}$, and which value of M$_{\rm BH, min}$ best reproduces the LFBOT rate? To explore this, we couple population synthesis with a model for the metallicity-dependent cosmic star formation rate history.

For the population synthesis we use the Binary Population and Spectral Synthesis ({\sc bpass}) models \citep[v2.2.1,][]{2017PASA...34...58E,2018MNRAS.479...75S}. These are a publicly available grid of pre-calculated binary stellar evolution models, in which the primary evolution is followed in detail, while the secondary is evolved with the rapid evolution prescriptions of \citet{2002MNRAS.329..897H}. Models are provided at 13 metallicities and at each metallicity, each stellar model is weighted according to observed binary parameter distributions \citep{2017ApJS..230...15M}. We adopt the model set with a \citet{2001MNRAS.322..231K} broken power-law initial mass function (IMF), with a minimum mass of 0.1\,M$_{\odot}$ and a maximum of 300\,M$_{\odot}$, a break mass of 0.5\,M$_{\odot}$ and a slope of $\alpha = -1.30$ ($-2.35$) below (above) the break. At each metallicity $Z$, the weighting of each model corresponds to the number of systems resembling that model expected in a stellar population of $10^{6}$\,M$_{\odot}$. For full details about the binary stellar models and the code, we refer the reader to the {\sc bpass} v2 release paper \citet{2017PASA...34...58E}.

For the metallicity-dependent cosmic star formation rate history, CSFH$(Z,z)$, we adopt the model of \citet{2006ApJ...638L..63L}. We seed the population synthesis models according to this CSFH such that the volumetric birth rate of each system ({\sc bpass} model) at each redshift $z$ and metallicity $Z$ is proportional to the product of the star formation rate density SFRD$(Z,z)$ and the model weighting (defined by the IMF and binary parameter distributions as described above). 

We adopt the standard {\sc bpass} assumptions that core-collapse occurs when, at the end of the model, the total mass exceeds 1.5\,M$_{\odot}$, the CO core mass exceeds 1.38\,M$_{\odot}$ and the ONe core mass is non-zero. Remnant masses are a pre-calculated output of the models. They are determined by injecting 10$^{51}$\,erg of energy, calculating the mass in the outer layers which can be lifted to infinity by this energy injection, and taking the remaining mass as the remnant mass \citep[for full details see][]{2004MNRAS.353...87E,2017PASA...34...58E}. We note that this energy injection is likely an overestimate for BH-forming events and means we have a large amount of mass potentially available for fallback accretion (this is addressed further in the discussion). We classify all core-collapse remnants with M$<$3\,M$_{\odot}$ as neutron stars, and those heavier as black holes. We assume that all NS-forming events produce successful supernovae, and that BH-forming events do not. Pair-instability supernovae (PISNe) are deemed to occur if the final carbon-oxygen core mass exceeds 60\,M$_{\odot}$ and the final helium core mass is $<$133\,M$_{\odot}$. In this case no remnant is left behind \citep{2002ApJ...567..532H}. Pulsational pair instability supernovae (PPISNe) are accounted for by adopting the remnant mass prescription of \citet{2019ApJ...887...53F} for final carbon-oxygen masses between 38 and 60\,M$_{\odot}$, following \citet{2023MNRAS.520.5724B}. The true landscape of explodability and remnant formation is more complex, and our assumptions neglect islands of explodability above the mass regime where neutron stars are formed, but capture the broad behaviour predicted by many works.\citep[e.g.][]{2009MNRAS.395.1409S,2009ARA&A..47...63S,2012ApJ...749...91F,2020ApJ...890...51E,2020MNRAS.492.2578S,2020MNRAS.499.2803P,2021ApJ...909..169K,2022MNRAS.511..903P,2025A&A...695A..71L,2025A&A...695A.122U,2025A&A...700A..20M}. These assumptions also naturally reproduce observed (binary) remnant mass distributions \citep[e.g.][]{2014ApJ...785...28K,2023MNRAS.520.5724B,2023A&A...676A..31D,2025arXiv250818083T}. Nevertheless, we acknowledge the possible impacts of these simplifying assumptions in Section \ref{sec:explode}.

The remnant formation time for each stellar model is the seed time, plus the age of the model at core-collapse (although for these short-lived stars, this is negligible compared with cosmological timescales). In this way, we construct a metallicity-weighted supernova rate and BH formation rate as a function of redshift. Since all (confirmed) LFBOTs thus far have redshifts $z<0.4$, we compare the mean rate of supernovae and BH-forming events over this redshift range, for different values of M$_{\rm BH, min}$. Specifically, we calculate the mean ratio of the supernova rate and the formation rate of BHs with mass M$_{\rm BH, min}$ or greater. As the form of CSFH$(Z,z)$ is uncertain, we repeat the process with two extreme variations of the CSFH$(Z,z)$. These are defined by shifting the \citet{2006ApJ...638L..63L} model by $\pm$0.2\,dex in 12$+$log(O/H), which captures the approximate envelope of possible CSFHs found by \citet{2019MNRAS.488.5300C}.

\begin{figure}
\centering
\includegraphics[width=0.99\columnwidth]{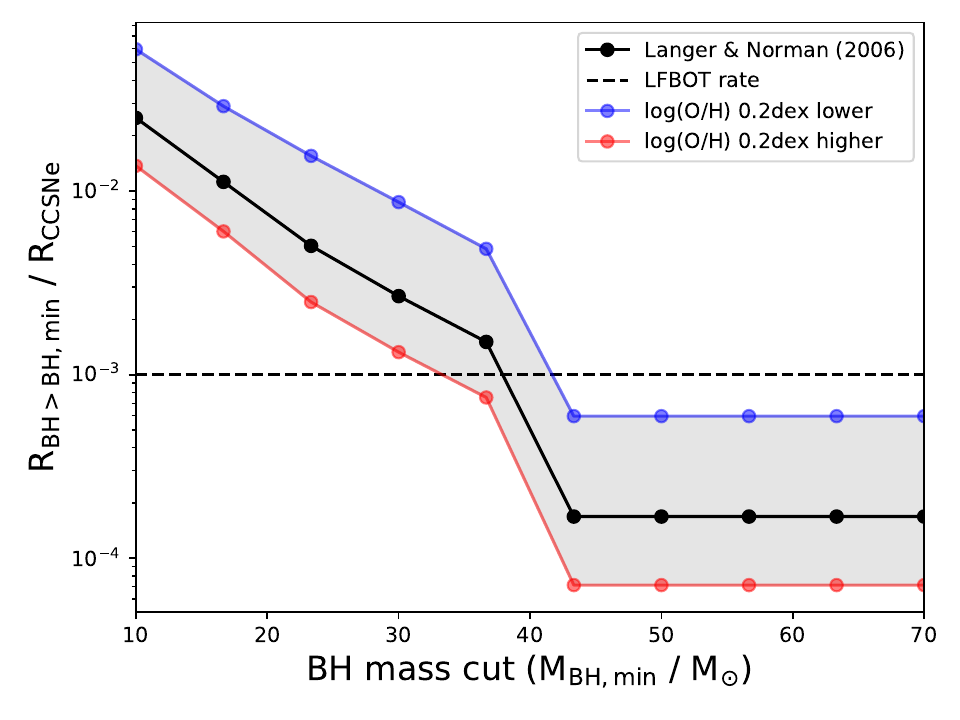}
\caption{The mean ratio of the rate of black hole formation above a minimum mass M$_{\rm BH,min}$ to the rate of core-collapse supernovae at z$<0.4$. The black line shows this ratio as a function of M$_{\rm BH,min}$ assuming the metallicity-dependence cosmic star-formation history (CSFH) of \citet{2006ApJ...638L..63L}. The red and blue lines, bounding the grey shaded region, are the result of adopting a CSFH shifted up/down by 0.2\,dex in 12$+$log(O/H) respectively. The dashed horizontal line is the estimated LFBOT volumetric rate \citep{2023ApJ...949..120H}. We therefore find that the predicted formation rate of BHs with masses greater than (38$^{+3}_{-5}$)\,M$_{\odot}$ is consistent with the LFBOT event rate.}
\label{fig:Rbh}
\end{figure}

The results are shown in Figure \ref{fig:Rbh}. Using the fiducial CSFH we find that the birth rate of BHs at $z<0.4$ with masses $\gtrsim$38\,M$_{\odot}$ is $\sim$0.1\% of the CCSN rate, matching LFBOTs. The progenitor temperatures, luminosities and radii pre-collapse are provided in Figure \ref{fig:HRD} of Appendix \ref{sec:apx}. For a CSFH of the same form but peaking at a metallicity 0.2\,dex lower in 12$+$log(O/H), we find a higher BH mass threshold of $\gtrsim$41\,M$_{\odot}$ is required, because there are more high-mass BHs being formed overall with respect to CCSNe. The reverse is true for a CSFH which peaks 0.2\,dex higher, yielding a threshold of $\gtrsim$33\,M$_{\odot}$. Therefore, varying the CSFH changes the absolute number of BHs being born \citep[although not the general shape of the BH mass distribution,][]{2023ApJ...948..105V}, and the number formed with respect to CCSNe. 

Above mass cuts of $\sim$45\,M$_{\odot}$, the ratio of birth rates remains constant, because the cut has moved into the PISN mass gap, and only black holes with masses above the gap are contributing. In summary, we find that the birth rate of BHs heavier than several tens of Solar masses is similar to the LFBOT rate, and that this result is robust against uncertainties in the CSFH.

\section{Metallicity dependence and LFBOT host galaxies}\label{sec:hosts}

\begin{table}
	\centering
	\caption{LFBOTs, their host metallicities and references.} 
	\label{tab:host_galaxy}
	\begin{tabular}{ccc} 
		\hline
		\hline
		  LFBOT  & Z/Z$_{\odot}$ & Reference \\
		\hline
            AT\,2018cow & 0.47$\pm$0.01 & \citet{2020MNRAS.495..992L}$^{a}$ \\ 
            ZTF18abvkwla & $\sim$0.37 & \citet{2020ApJ...895...49H} \\
            CSS161010 & $\sim$0.15 & \citet{2024ApJ...977..162G} \\
            ZTF20acigmel & 0.15$^{+0.15}_{-0.07}$ & \citet{2021MNRAS.508.5138P}$^{b}$ \\
            AT2020mrf & 0.27$^{+0.07}_{-0.05}$ & \citet{2022ApJ...934..104Y} \\
            AT2023fhn & 0.08$\pm$0.02 & \citet{2024AandA...691A.329C} \\
            AT2024wpp & 0.53$^{+0.08}_{-0.12}$ & \citet{2025MNRAS.537.3298P}$^{b}$ \\ 
		\hline
	\end{tabular}
    \tablefoot{
Other  LFBOT candidates exist, but have insufficient follow up to confirm their nature or to characterise the host galaxy (e.g. we have no $Z$ estimate for AT2022tsd). Where metallicities are reported in 12+log(O/H), we convert to mass fractions following a linear interpolation of the values in table 2 of \citet{2018MNRAS.477..904X}. For CSS161010, we adopt the gas-phase metallicity determined with emission lines \citep{2024ApJ...977..162G}, as this is likely more representative of recently formed stars in the galaxy.\\
\tablefoottext{a}{IFU metallicity measurement, resolves the host giving a local (few hundred parsec scale) value.}
\tablefoottext{b}{Metallicity derived from the host's total stellar mass using equation 3 of \citet{2004ApJ...613..898T}. Uncertainties are solely propagated from uncertainties on the galaxy stellar mass, and do not include scatter in the $M-Z$ relation.}
}
\end{table}

Given a black hole mass cut M$_{\rm BH, min}$ which yields a BH formation rate matching the observed LFBOT rate, for an assumed CSFH, we can next ask: what is the (CSFH weighted) metallicity distribution of the progenitors of these black holes, and how does this compare with the host galaxies of LFBOTs? We collate the available information in the literature on LFBOT host metallicities in Table~\ref{tab:host_galaxy}. Four of the seven have host-averaged metallicities derived from spectral energy distribution (SED) fitting. In one case (AT\,2018cow) there is integral field unit data available, enabling a more local \citep[resolution of a few hundred parsec,][]{2020MNRAS.495..992L} measurement of the metallicity. For the remaining two, no direct measurement has been published, so we obtain an approximate value through the galaxy mass-metallicity relation \citep{2004ApJ...613..898T}. We adopt Z$_{\odot}=0.02$ by mass fraction for Solar metallicity, but note that this specific choice has no impact on the comparisons or analysis since both the observed and synthetic metallicities are scaled by the same factor.

\begin{figure}
\centering
\includegraphics[width=0.99\columnwidth]{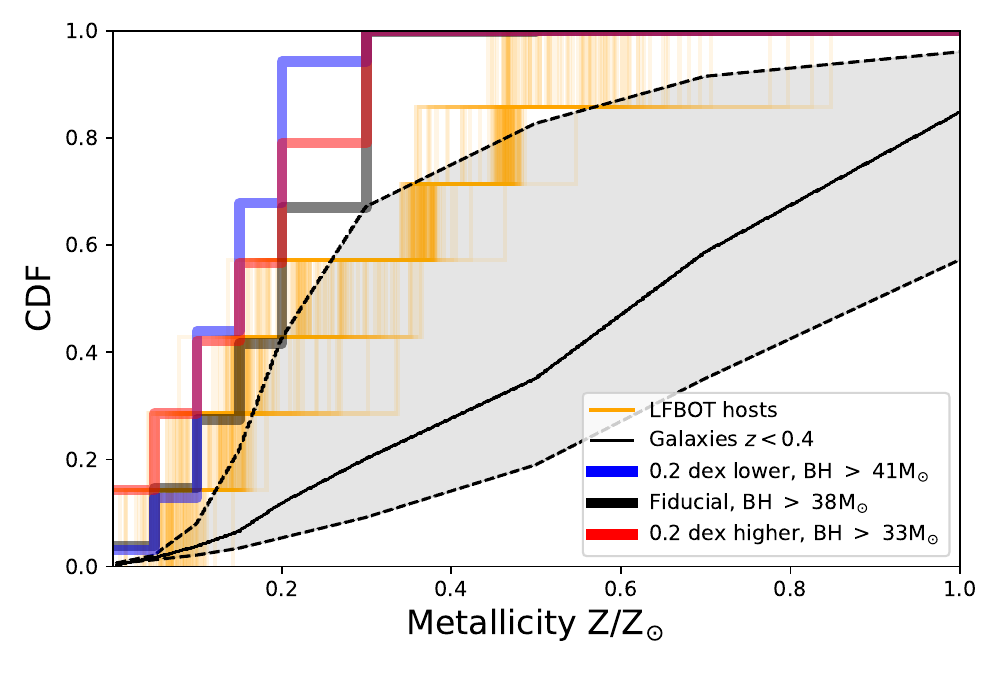}
\caption{The cumulative distribution of LFBOT host galaxy metallicities is shown in orange (see Table \ref{tab:host_galaxy}). We sample the Z values, assuming Gaussian uncertainties, 100 times, producing the many realisations of the cumulative distribution shown. Cumulative distributions of our selected progenitor metallicities, weighted by the CSFH(Z) at z$<$0.4, are also shown. The black line is the $z<0.4$ mean metallicity distribution of star-forming galaxies \citet{2006ApJ...638L..63L}. The grey shaded region bounded by dashed lines is defined by shifting the distribution of \citet{2006ApJ...638L..63L} by $\pm$0.2 in 12+log$_{10}$(OH), covering the high and low Z extremes defined by \citet{2019MNRAS.488.5300C}.}
\label{fig:Z}
\end{figure}

In Figure~\ref{fig:Z} we compare the number of black holes more massive than M$_{\rm BH,min}$ born per $10^{6}$\,M$_{\odot}$ of star formation, weighted by the mean $z<0.4$ metallicity distribution for each of the three CSFHs adopted, with the measured distribution of LFBOT host or environmental metallicities. There is a strong preference for BHs with masses greater than 38\,M$_{\odot}$ to be born in stellar populations with metallicity less than $\sim$0.3\,Z$_{\odot}$. Below this threshold, the distributions are relatively constant: this reflects that the fact that (i) we are tailoring our mass cut to yield a BH formation rate that is 0.1\% of the CCSN rate for each CSFH, and (ii) that the CCSN rate itself is not substantially varying with the CSFH variations. This is because NS formation efficiency, unlike BH formation efficiency, does not sharply decline at higher metallicities \citep[e.g.][]{2025ApJ...979..209V}.

It is immediately apparent that LFBOT environments sample the Z$ \lesssim 0.5$Z$_{\odot}$ region. The formation of BHs with M$>$M$_{\rm BH,min}$ across this range instead samples metallicities Z$<0.3$Z$_{\odot}$. A KS-test between the LFBOT host values and the fiducial $>38$\,M$_{\odot}$ $Z$ distribution returns $p=0.01$, formally rejecting the null hypothesis that the observed host values are consistent with the synthetic distribution. However, we note that the host metallicities are far from uniformly determined - with methods ranging from resolved IFU spectroscopy, probing the local environment within the host, to host-integrated metallicites to the mass-metallicity relation. This likely introduces errors/scatter which has not been accounted for, and host-integrated measurements will naturally be higher than the lowest metallicity environments within each host. There may also be pockets of low metallicity gas within and around star-forming galaxies, driving low-Z star-formation \citep[e.g.][]{2015A&A...582A..78M,2023MNRAS.520..879M}. Nevertheless, the current information suggesting that LFBOTs can occur in environments with metallicities above the strong cut-off predicted is a challenge to the model. This is most notable for AT2018cow, whose metallicity measurement is integrated across the immediate (few hundred pc) environment \citep{2020MNRAS.495..992L}, but even in this case variations on smaller scales are possible.

\section{Expectations for the emission}\label{sec:emiss}
\subsection{H and He-poor spectra}
\begin{figure}
\centering
\includegraphics[width=0.99\columnwidth]{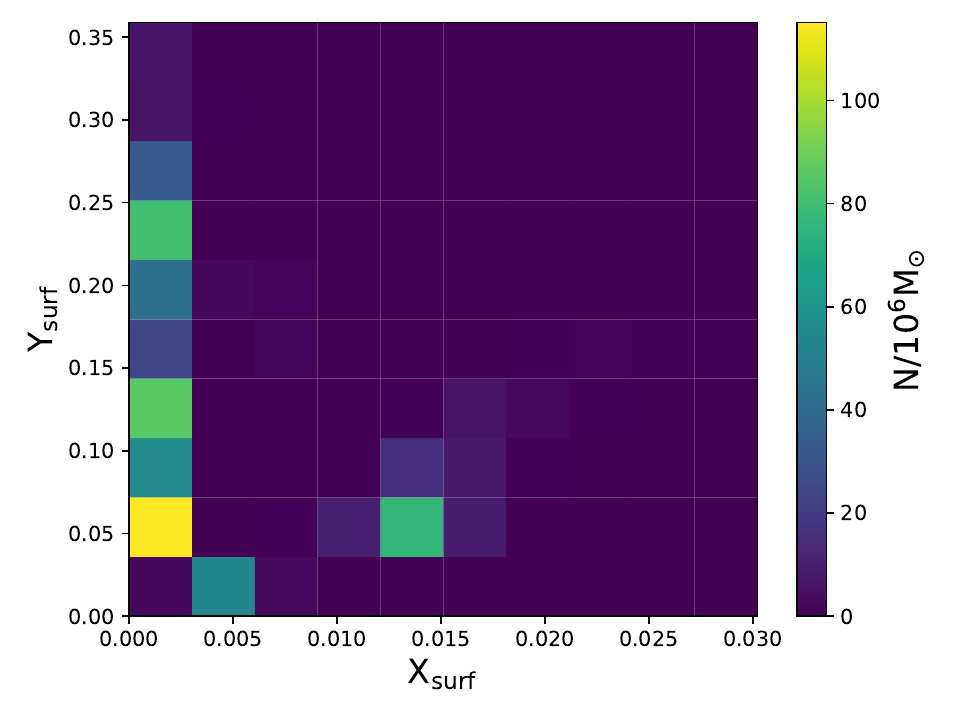}
\caption{The leftover (i.e. ejected and/or accreted) mass fractions of hydrogen (X$_{\rm surf}$) and helium (Y$_{\rm surf}$) in the last time-step of the selected {\sc bpass} models, with contributions from different metallicities determined by the mean metallicity spread at $z<0.4$ \citep{2006ApJ...638L..63L}. The colourbar indicates the number of events per 10$^{6}$\,M$_{\odot}$ of star formation, with contributions from different metallicities following the fiducial case as shown in Figure \ref{fig:Z}, and model weightings as defined in Section \ref{sec:rates}.}
\label{fig:XY}
\end{figure}

LFBOTs display H and He poor spectra at early times \citep[e.g.][]{2018ApJ...865L...3P}. The progenitor star, or star being tidally disrupted, therefore must have a He or He-poor composition. In the TDE scenario, a white dwarf is favoured \citep{2019MNRAS.487.2505K}. For a BH-stellar merger, the stellar object must be a Wolf-Rayet star \citep{2022ApJ...932...84M}. In the scenario explored in this paper, the median H and He mass fractions in the `leftover mass'\footnote{I.e. the stellar mass leftover after remnant formation. In a successful supernova with no fallback, this would be the ejecta mass.} of the models selected as progenitors of BHs with $M>38$\,M$_{\odot}$ are X$_{\rm surf} = $0.003$^{+0.005}_{-0.003}$ and Y$_{\rm surf} = $0.08$^{+0.12}_{-0.05}$ respectively (where the uncertainties are defined by the 16$^{\rm th}$ and 84$^{\rm th}$ percentiles). The distributions of X$_{\rm surf}$ and Y$_{\rm surf}$ are shown in Figure \ref{fig:XY}. 69\% of our selected models are classified as Wolf-Rayet stars given the X$_{\rm surf}$ values and high surface temperatures of log$_{10}$(T/K)$>$4.45 \citep{2017PASA...34...58E}. The outer layers of these stars therefore meet the criteria for being H and He poor, and successfully exploding stars with such low X and Y surface mass fractions are expected to produce type Ic supernovae \citep[i.e. SNe without detectable H and He lines, e.g.][]{2012MNRAS.424.2139D,2013MNRAS.436..774E,2020MNRAS.491.3479C}.

\subsection{Dense circumstellar media and radio emission}
\begin{figure}
\centering
\includegraphics[width=0.99\columnwidth]{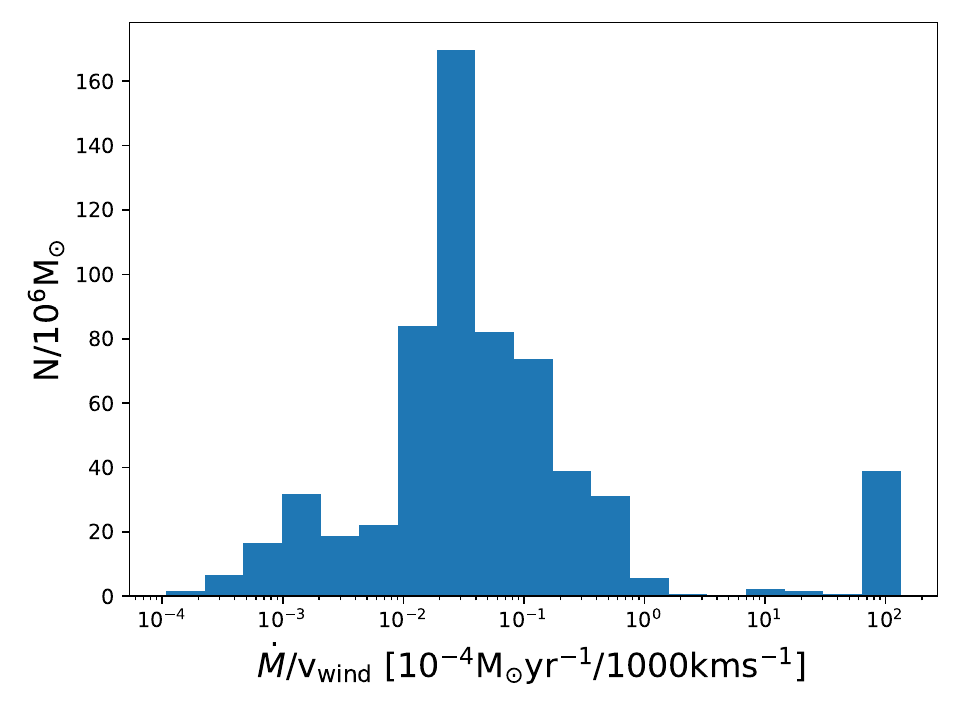}
\caption{The wind density parameter for the selected models, normalised by a typical Wolf-Rayet mass-loss rate and wind speed. The circumstellar densities resulting from these values are lower than typical Wolf-Rayet stars and lower than observed in some LFBOTs, likely due to weaker winds as a result of the strong low-metallicity bias introduced by requiring such massive black holes. Typical LFBOT values are in the range 1--100 \citep[e.g.][]{2022ApJ...932..116H,2022ApJ...926..112B}. The $y$-axis indicates the number events per 10$^{6}$\,M$_{\odot}$ of star formation, as described in Figure \ref{fig:XY}. }
\label{fig:csm}
\end{figure}

A key characteristic of LFBOTs is their slow-rising, luminous radio emission, attributed to self-absorbed synchrotron emission from a blast-wave expanding through the circumstellar medium (CSM). These blast-waves are mildly relativistic \citep[e.g.][]{2020ApJ...895L..23C}, and probe the CSM at radii of $\sim$10$^{16}$--10$^{17}$\,cm. LFBOT radio emission is well described by blast-wave propagation through dense CSM profiles defined by $\dot{M}/V_{\rm w}$ values of $\sim$1--100, in units of $10^{-4}$M$_{\odot}$\,yr$^{-1}$/1000\,km\,s$^{-1}$. For Wolf-Rayet wind speeds of order 1000\,km\,s$^{-1}$, radii of  $\sim$10$^{16}$--10$^{17}$\,cm corresponds to mass loss in the final decades before explosion. We calculate the mean final $\dot{M}/V_{\rm w}$ values for the selected models, noting that {\sc bpass} only runs until the end of core carbon burning. Mass loss rates are provided as a {\sc bpass} output. We adopt standard {\sc bpass} wind speed prescriptions, namely those of \citet{2001A&A...369..574V}, and \citet{2000A&A...360..227N} for Wolf-Rayets \citep[we refer the reader to][for more details]{2017PASA...34...58E,2022MNRAS.515.2591C}. We find a median final $\dot{M}/V_{\rm w}$ = (0.04$^{+0.21}_{-0.03}$)\,$10^{-4}$M$_{\odot}$\,yr$^{-1}$/1000\,km\,s$^{-1}$. The full distribution of $\dot{M}/V_{\rm w}$ is shown in Figure \ref{fig:csm}. The bulk of our progenitor mass-loss rate and wind speed values lie below the values inferred from LFBOT radio observations, which span the range $\sim$1-100 in these units \citep[e.g.][under the assumption of $\epsilon_{e}=0.1$ and $\epsilon_{B}=0.01$]{2022ApJ...932..116H,2022ApJ...926..112B}. We note that mass loss in the final decades before core-collapse may be significant and eruptive, plausibly explaining the non-$r^{-2}$ profiles observed \citep[e.g.][]{2014ARA&A..52..487S}. However, because the stellar evolution models used end after core carbon burning, a full modelling of the CSM around these stars on the relevant scales is beyond the scope of this paper.

\subsection{Long-lived ultraviolet plateaus}\label{sec:plateau}
\begin{figure}
\centering
\includegraphics[width=0.99\columnwidth]{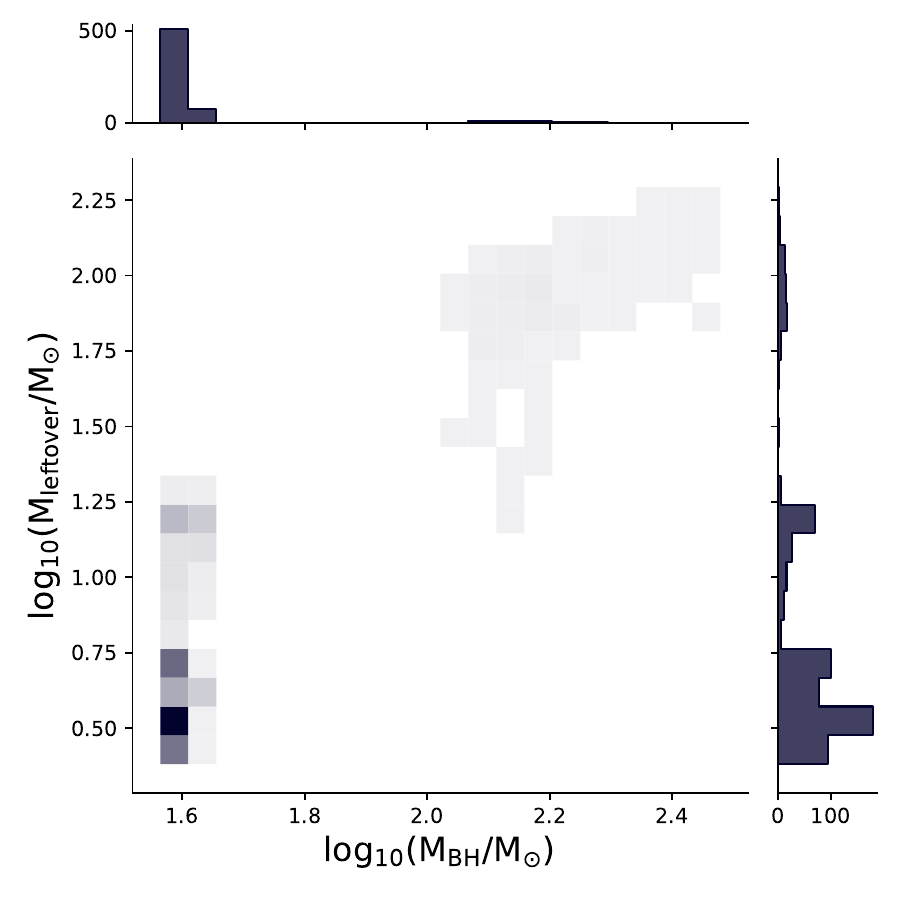}
\caption{Remnant black hole mass versus the `leftover' mass for the models selected when the fiducial CSFH is used (black lines in Figures \ref{fig:Rbh} and \ref{fig:Z}). These are the remnant and leftover (ejecta) masses when an core-collapse energy injection of 10$^{51}$\,erg is assumed \citep{2004MNRAS.353...87E}, taking into account the mass gap from PISNe and PPISNe \citep{2019ApJ...887...53F,2023MNRAS.520.5724B}. The histograms show the number events expected per 10$^{6}$\,M$_{\odot}$ of star formation, as described in Figure \ref{fig:XY}.}
\label{fig:accrete}
\end{figure}

Finally, we ask whether the selected stellar models can produce a long-lived tail of UV emission lasting at least several years after the event, as observed in AT2108cow \citep{2022MNRAS.512L..66S,2023MNRAS.519.3785S,2023ApJ...955...43C,2023MNRAS.525.4042I,2025MNRAS.544L.108I}. This has been interpreted as due to a long-lived accretion disc, reminiscent of the emission seen in TDEs \citep{2022ApJ...932...84M,2023MNRAS.525.4042I,2025MNRAS.541..429M}. To see whether our models can produce AT\,2018cow-like UV plateaus through this mechanism, we first examine the remaining mass in the selected {\sc bpass} models which survives outside the nascent BHs (i.e., the pre-core-collapse mass minus the initial BH mass). In Figure \ref{fig:accrete} we show 2D histograms of these masses at each metallicity, for the fiducial \citet{2006ApJ...638L..63L} CSFH. For more details on the BH mass distribution in {\sc bpass}, we refer the reader to \citet{2017PASA...34...58E}, \citet{2022MNRAS.511.1201G} and \citet{2023MNRAS.520.5724B}. More detail on the progenitors, in terms of temperature, luminosity and radius at the end of the models (i.e. the end of core carbon burning) is provided in Section \ref{sec:apx} of the Appendix.

As outlined in Section \ref{sec:rates}, {\sc bpass} calculates BH masses by injecting a standard supernova energy of 10$^{51}$\,erg to determine the mass of the material which is unbound. The remainder is the remnant mass. In a failed supernova, the `leftover' material is not successfully ejected. It may therefore be reasonable to assume that all of the leftover material falls back and is accreted instead, in line with other prescriptions for final BH masses in the massive to very massive star regime, which have close to a one-to-one relation between the pre-core-collapse mass and final remnant mass (after fallback) for stripped envelope progenitors \citep{2012ApJ...749...91F,2023MNRAS.520.5724B}. With the BH mass M$_{\rm BH}$, leftover mass  M$_{\rm leftover}$, total pre-collapse stellar mass M$_{\star}$, radius R$_{\star}$ and rotational velocity v$_{\star}$ as inputs, we determine an initial disc radius $r_{0}$ as follows,
\begin{equation}\label{eq:1}
    r_{0} \approx 0.01 f_{\rm j,disc}^{2} R_{\star} \bigg(\frac{M_{\star}}{f_{\rm acc} M_{\rm leftover}}\bigg)^{2} \bigg(\frac{M_{\star}}{M_{\rm BH}}\bigg)
\end{equation}
where $f_{\rm j,disc}$ and $f_{\rm acc}$ are the fraction of the pre-collapse angular momentum and leftover mass which go into the disc, respectively. A derivation of Equation \ref{eq:1} is provided in Appendix \ref{sec:apx2}. Since stellar rotation is not explicitly tracked in {\sc bpass}, we have assumed the pre-collapse rotational velocity to be 10\% of the breakup (critical) velocity \citep[massive stars in low metallicity environments such as the Magellanic Clouds are observed to be rotating at $\sim$10--20\% of their critical velocity e.g.][]{2013A&A...560A..29R,2015A&A...580A..92R}. We proceed to use the model of \citet{2020MNRAS.492.5655M,2025MNRAS.544.2225M} to calculate UV light-curves for every selected stellar model. The final parameter required (beyond the disc mass, black hole mass and radial scale) is the initial evolutionary timescale of the disc. We follow standard \cite{SS73} theory and take $t_{\rm visc} = \alpha^{-1} \theta^{-2} \sqrt{r_0^3 / GM_{\rm BH}}$, where $\alpha \sim 0.1$ is a dimensionless factor, and $\theta \equiv h/r$ is the disc aspect ratio. Anticipating a thick disc (for this likely initially super-Eddington flow) we sample $\alpha^{-1}\theta^{-2}$ in the range $\sim (50, 500)$, inducing scatter in our simulated light curves. This fixes all parameters in the disc model. 

\begin{figure*}
\centering
\includegraphics[width=0.99\textwidth]{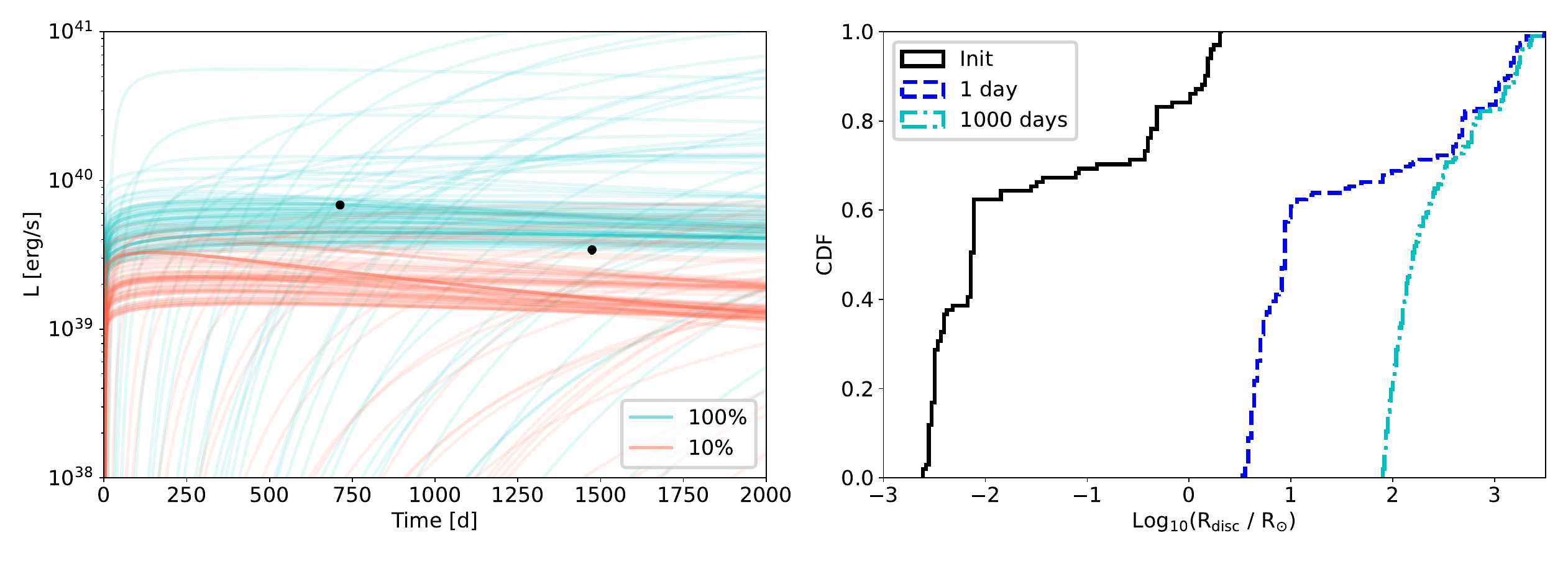}
\caption{Left: Predictions for UV light-curves arising from the accretion disc around natal BHs of mass greater than 38\,M$_{\odot}$ in our failed SN models, using the model of \citet{2020MNRAS.492.5655M,2025MNRAS.544.2225M}. These predictions are for observations in the {\it Hubble Space Telescope} WFC3/F225W filter; the black points are late-time F225W observations of AT\,2018cow \citep{2023MNRAS.525.4042I}. Two example cases are shown: one assuming that 10\% of the total pre-collapse angular momentum, and 10\% of the leftover mass, goes into the disc (red). The other is the extreme case that 100\% of the available angular momentum and mass goes into the disc (cyan). Both scenarios adopt progenitor rotation at 10\% of the critical velocity. The large spread in luminosity in both the cyan and red curves is due to the spread in black hole and leftover masses (spanning the pair instability mass gap, see Figure~\ref{fig:accrete}). Right: disc radii for the fiducial selected models at $t=0$, 1 and 1000 days. The radii are the same for both the 10\% and 100\% assumptions (see Equation \ref{eq:1}). The discs spread rapidly, with around 40\% reaching radii of 30--40\,$R_{\odot}$ within the first day (the scale inferred from blackbody modelling of LFBOT observations). All discs ultimately reach 100--1000\,$R_{\odot}$ by 1000 days.} 
\label{fig:LCs}
\end{figure*}

We adopt two example scenarios, and show the predicted UV lightcurves in Figure \ref{fig:LCs} for the F225W Wide Field Camera 3 (WFC3) {\it Hubble Space Telescope} filter. In first two, only 10\% of the available angular momentum and mass is transferred to the disc. In the second extreme case, all of the angular momentum and mass goes into the disc. We acknowledge that the details details of supernova fallback, such as its asphericity and the fraction which falls back versus being ejected, are not captured in the models or our simple parametrisation. This motivated our choice of two scenarios which nevertheless represent a wide range of possible fallback masses and angular momenta, including the hypothetical maximum case. The total leftover mass is also uncertain (and depends on the injected energy at core-collapse), for a discussion of the impact of this we refer the reader to Section \ref{sec:explode}. 

The resulting light-curves can only match the observed late-time F225W flux for AT\,2018cow if a large fraction of the leftover mass and angular momentum are transferred to the disc (see also Omand et al. in prep). The initial disc radii predicted from these models are considerably smaller than the $\sim$30-40\,R$_{\odot}$ inferred from the late-time fits to AT2018cow \citep{2023MNRAS.525.4042I,2024ApJ...963L..24M}. This is because the majority of the progenitors are compact stars (e.g. Wolf-Rayets, see Figure \ref{fig:HRD} of Appendix \ref{sec:apx}). However, the discs spread rapidly (as t$^{2/3}$), reaching $\sim$10\,R$_{\odot}$ within a day and  $\sim$100\,R$_{\odot}$ within 1000 days. We note that the factors f$_{\rm acc}$ and f$_{\rm j,disc}$ cancel, such that the two cases adopted return the same initial disc radii. The growth of the outer disc radius will be less pronounced if angular momentum is transported away from the disc by winds (potentially likely at early times), see Appendix \ref{sec:apx2}. We find that $t_{\rm visc}$ covers a wide range of $\sim$0.5-10$^{6}$ days. Only for the shortest of these timescales can the same accretion mechanism plausibly explain the rapid early-time rise in LFBOT lightcurves. Otherwise, we acknowledge that some other process must drive the initial rise and peak.

The failed SN interpretation of LFBOTs is predicated on the notion that some non-negligible fraction of the total stellar mass does not participate in the prompt BH formation and instead falls back onto the nascent BH. If the core-collapse of such massive stars actually results in the entire stellar mass being promptly engulfed by the event horizon, there will no material to accrete, ruling out failed supernovae as a viable channel for LFBOTs. However, in our fiducial models this is not the case (see Figure~\ref{fig:accrete}, as well as a discussion of the explodability assumptions in Section \ref{sec:explode}).

\section{Discussion}\label{sec:disc}
\subsection{LFBOT environments}
In the failed SN scenario, we would expect LFBOTs to be strongly associated with star-forming regions, as the short-lived massive progenitors are not expected to travel far from their birth sites. Although the LFBOT sample is small, it appears as though LFBOTs occupy diverse environments within their hosts, with at least two events at moderate to large host-normalised offsets \citep[][]{2024MNRAS.527L..47C,2025MNRAS.537.3298P}. The apparent preference for LFBOTs to occur in low metallicity environments may help in this regard: in general, galactic outskirts have lower metallicity than their inner regions \citep[e.g.][]{2012A&A...540A..56P,2025ApJ...978L..39J}. However, a similar argument could be made for e.g. core-collapse gamma-ray bursts, which are typically not observed at large offsets. Regardless of offsets, the host metallicities of LFBOTs extend above the sharp cut-off predicted for very massive star core-collapse events (see Sec \ref{sec:hosts}). Furthermore, metallicities from SED fitting should be treated with caution \citep[e.g.][]{2025A&A...695A..86N}. For example, the mass-metallicity relation for the host of AT\,2023fhn suggests a metallicity of $\sim$0.5\,Z$_{\odot}$, versus the fitted value of $\sim$0.1\,Z$_{\odot}$. In any case, the metallicity distributions as presented in Fig.~\ref{fig:Z} are inconsistent, but the possibility of low-Z pockets on spatial scales smaller than probed by current measurements has the potential to resolve this discrepancy.

\subsection{Initial mass function}
We have adopted model weightings with an initial mass function (IMF) slope above 0.5\,M$_{\odot}$ of -2.35. Two other high-end slopes are available with the {\sc bpass} outputs: -2.00 and -2.70. If the -2.70 slope is adopted, fewer high-mass black holes are born, and the cut-off M$_{\rm BH,min}$ drops to $\sim$30\,M$_{\odot}$ such that the BH formation rate remains consistent with the LFBOT rate (0.1\% of the CCSN rate). The number of neutron star progenitors also decreases, such that the SN rate decreases, reducing the amount by which M$_{\rm BH,min}$ has to decrease in order to maintain 0.1\% of the CCSN rate. If -2.00 is adopted, there are more black holes and the cut-off moves up to $\sim$40\,M$_{\odot}$, around the lower edge of the PISN mass gap. This range (i.e. $\sim$30--40\,M$_{\odot}$) represents the uncertainty in the minimum black hole mass required to match the LFBOT rate which arises from the choice of IMF, and is slightly larger than the uncertainty arising solely from the choice of CSFH$(Z,z)$. Nevertheless, the broad conclusion is that the formation rate of black holes with masses greater than $\sim$30--40\,M$_{\odot}$ is consistent with the LFBOT rate.

\subsection{Explodability and remnant mass assumptions}\label{sec:explode}
The results in this paper depend on the BH masses formed and the mass available in the `leftover' material for fallback. The models adopted calculated BH masses by injecting 10$^{51}$\,erg of energy and determining the mass which which remains bound. If the actual energy injection in failed events is $<<$10$^{51}$\,erg \citep[as expected, e.g.][]{2018MNRAS.476.2366F,2023MNRAS.526..152K} then the BH masses will be greater. However, the ejecta/leftover masses are typically not more than 10\% of the BH mass, so the most they could increase by is $\sim$10, which would not significantly change our conclusions. Ultimately, the models predict similar final remnant masses to other widely used prescriptions in the high mass regime \citep[e.g.][see also Figure \ref{fig:core} of Appendix A]{2012ApJ...749...91F}.

There is also the question of which stars explode and which undergo a failed collapse. Our assumption that all BH-forming events fail is an upper limit on the rate of failed SNe. The predicted rate of all failed SNe under this assumption is $\sim$0.2 of the CCSN (NS) formation rate, which is compatible with upper limits on the rate of failed SNe from searches for faint, long-lived red transients \citep[thought to be a likely outcome for failed SNe in the absence of any additional energy injection from a disc,][]{2022MNRAS.514.1188B}.  

However, we are likely overestimating the failed SN rate, given recent hydrodynamical simulations and suggestions that there may not be a missing red supergiant problem \citep[e.g.][]{2025ApJ...987..164B,2025ApJ...979..117B}. If failed SNe only occur in the 12-15\,M$_\odot$ (ZAMS mass) window where failed events are reliably predicted to occur in these simulation (and assuming NS formation everywhere outside this range unless a PISN occurs), then we would expect a ratio of failed events to CCSNe of $\sim$0.15. The lower masses (higher up the IMF) counteract the narrower mass range, producing a similar ratio to our fiducial $>$30--40\,M$_{\odot}$ BH mass cut. Such a scenario could therefore reproduce the LFBOT rate but not the inferred BH masses. The prospect of incorporating more physically motivated prescriptions for remnant and ejecta mass distributions in the framework of population synthesis \citep[e.g.][]{2020MNRAS.499.2803P,2025A&A...700A..20M} could help refine results such as these in future.

\subsection{A potential source of r-process nucleosynthesis}
Massive black holes and their accretion discs, such as those discussed here, have also been predicted as possible sites of `super-kilonovae'. In this scenario, the density in a high $\dot{M}$ accretion disc becomes high enough to trigger r-process nucleosynthesis \citep{2019Natur.569..241S,2022ApJ...941..100S,2024PhRvD.110h3024D,2024ApJ...968...14R,2025arXiv250315729A}. The possibility of such a channel has implications for where and when r-process elements are produced in the Universe \citep{2025ApJ...982..144N}. Given the black hole and accreted masses involved in the failed supernova interpretation of LFBOTs, they may be promising sites for r-process production. The volumetric rate of LFBOTs in the redshift range considered in this work \citep[0.1\% of the CCSN rate, or 70\,yr$^{-1}$\,Gpc$^{-3}$]{2023ApJ...949..120H} is compatible with the rate of super-kilonovae predicted by \citet{2022ApJ...941..100S} of 10--100\,yr$^{-1}$\,Gpc$^{-3}$. We have shown that the failed-supernova interpretation of LFBOTs naturally favours events forming black holes greater than $30-40$\,M$_{\odot}$, while super-KNe have been predicted in events producing black holes greater than $\sim60$\,M$_{\odot}$ \citep[][]{2022ApJ...941..100S}. If the true rate of LFBOTs is just a factor of few lower than the current $\lesssim$70\,yr$^{-1}$\,Gpc$^{-3}$ best estimate, then the BH mass cut required to match the LFBOT rate moves up to the pair instability mass gap (see Figure \ref{fig:Rbh}), and we have a 1:1 mapping between LFBOT and super-KN progenitors. In any case, there is surely substantial overlap in the populations. Furthermore, a near-infrared excess - a hallmark of radiative reprocessing by r-process elements - was observed on a timescale of weeks following both AT\,2018cow \citep{2019MNRAS.484.1031P} and AT2024wpp \citep{2025MNRAS.537.3298P}. This has been described either as a dust echo or due to free-free opacity effects outside the optical photosphere in both events \citep{2023ApJ...944...74M,2025ApJ...993L...6N}, but no spectroscopic characterisation of this feature in an LFBOT has yet been obtained. Future (James Webb Space Telescope) near-infrared spectroscopic observations could potentially distinguish between an r-process element-forming super-kilonova and a featureless dust echo or free-free emission. However, in order for a UV-emitting disc and IR-emitting super-kilonova to be simultaneously observable, an anisotropic geometry which allows simultaneous viewing of both photospheres would be required.

\subsection{Further implications}
There are several other observations associated with LFBOTs which any model invoked to explain them must satisfy, assuming that LFBOTs are homogenous class. For example, extremely high velocity ($v=0.1c$), blue-shifted hydrogen lines were observed in the CSS161010 \citep[][]{2024ApJ...977..162G}, perhaps most easily explained as out-flowing streams of material in a tidal disruption event, although these have only been observed in CSS161010 thus far. Polarimetric observations have found that LFBOTs start highly polarised and trend towards low polarisation on a timescale of days \citep{2023MNRAS.521.3323M,2025MNRAS.537.3298P}. This can plausibly be explained in a failed supernova context by (i) an accretion disc producing the high early-time polarisation and (ii) a rapid transition to an spherical outflow \citep{2025MNRAS.537.3298P}, perhaps caused by outflows from highly super-Eddington accretion \citep[e.g.][]{2020ApJ...895L..23C}. The origin of the giant optical flares observed from AT2022tsd in the months after the event is unclear, but timescale arguments constrain the scale of the emitting region to the outer regions of an accretion disc, for a stellar to intermediate mass black hole central engine \citep{2023Natur.623..927H}.

\section{Conclusions}\label{sec:conc}
In this paper, we have investigated whether failed supernova in (very) massive star core-collapse events are plausible progenitors for Luminous Fast Blue Optical Transients (LFBOTs). Adopting a reasonable spread for the metallicity-dependent cosmic star formation history, we explored the rates, metallicities and expected observational characteristics of these events by identifying suitable progenitors in binary population synthesis models. Our main conclusion are as follows,
   \begin{enumerate}
      \item The rate of core-collapse events producing black holes more massive than $\sim$30--40\,M$_{\odot}$ is similar to the observed LFBOT rate,
      \item The expected metallicities of the progenitors of these black holes are comparable, albeit slightly lower and formally inconsistent, with the observed metallicities of LFBOT host galaxies. However, for most LFBOTs the measured metallicity is that of the host galaxy as a whole. Even for AT2018cow, whose metallicity is most discrepant with the predicted distribution, the spatial resolution of this measurement allows for the possibility of lower metallicity pockets on smaller spatial scales,
      \item Given that the progenitors of BHs with mass greater than $\sim$30--40\,M$_{\odot}$ have envelopes which are H and He poor, this is consistent with the early-time spectra of LFBOTs,
      \item We show that the mass loss rates and stellar wind speeds associated with these stellar models correspond to dense circumstellar media, but which nevertheless only overlap with the lower end of LFBOT radio observations, and struggle to reproduce the typical circumstellar densities inferred. However, as {\sc bpass} only runs until the end of core carbon burning, possible large mass loss phases occurring after carbon burning could explain the high-densities,
      \item The long-lived UV emission seen in AT\,2018cow can be reproduced in the failed supernova interpretation given black hole masses $\gtrsim 30$\,M$_{\odot}$, but this requires (i) several solar masses of fallback accretion and (ii) a high fraction of the pre-collapse stellar angular momentum to be transferred to the disc,
      \item The failed supernova interpretation of LFBOTs, and models for super-kilonovae from massive collapsars, share similar progenitors and black hole masses. Therefore, if LFBOTs are massive star core-collapse events, they may constitute a non-negligible contribution to the r-process budget of galaxies.
   \end{enumerate}
We therefore conclude that the failed supernova scenario is a plausible explanation for at least some LFBOTs, but it hinges on the measured metallicities at the LFBOT site and the origin of the dense circumstellar media. The true feasibility of this progenitor channel clearly depends on the detailed physics of very massive star core-collapse, late-stage mass loss, and the amount of mass which falls back or is ejected.

\begin{acknowledgements}
      We thank the referee Brian Metzger for their insightful comments on this manuscript.
      
      AAC acknowledges support through the European Space Agency (ESA) research fellowship programme. P.G.J. ~is supported  by the European Union (ERC, Starstruck, 101095973, PI Jonker). Views and opinions expressed are however those of the author(s) only and do not necessarily reflect those of the European Union or the European Research Council Executive Agency. Neither the European Union nor the granting authority can be held responsible for them.\\

      This work made use of v2.2.1 of the Binary Population and Spectral Synthesis ({\sc bpass}) models as described in \citet{2017PASA...34...58E} and \citet{2018MNRAS.479...75S}. This work has made use of {\sc ipython} \citep{2007CSE.....9c..21P}, {\sc numpy} \citep{2020Natur.585..357H}, {\sc scipy} \citep{2020NatMe..17..261V}; {\sc matplotlib} \citep{2007CSE.....9...90H}, Seaborn packages \citep{Waskom2021} and {\sc astropy},(\url{https://www.astropy.org}) a community-developed core Python package for Astronomy \citep{astropy:2013, astropy:2018}.
      
\end{acknowledgements}


%
\bibliographystyle{aa} 
\bibliography{aa57545-25.bib} 

@ARTICLE{Pringle81,
       author = {{Pringle}, J.~E.},
        title = "{Accretion discs in astrophysics}",
      journal = {\araa},
         year = 1981,
        month = jan,
       volume = {19},
        pages = {137-162},
          doi = {10.1146/annurev.aa.19.090181.001033},
       adsurl = {https://ui.adsabs.harvard.edu/abs/1981ARA&A..19..137P}
}

@ARTICLE{SS73,
       author = {{Shakura}, N.~I. and {Sunyaev}, R.~A.},
        title = "{Black holes in binary systems. Observational appearance.}",
      journal = {\aap},
         year = 1973,
        month = jan,
       volume = {24},
        pages = {337-355},
       adsurl = {https://ui.adsabs.harvard.edu/abs/1973A&A....24..337S}
}

@ARTICLE{2025MNRAS.537.3298P,
       author = {{Pursiainen}, M. and {Killestein}, T.~L. and {Kuncarayakti}, H. and {Charalampopoulos}, P. and {Warwick}, B. and {Lyman}, J. and {Kotak}, R. and {Leloudas}, G. and {Coppejans}, D. and {Kravtsov}, T. and {Maeda}, K. and {Nagao}, T. and {Taguchi}, K. and {Ackley}, K. and {Dhillon}, V.~S. and {Galloway}, D.~K. and {Kumar}, A. and {O'Neill}, D. and {Ramsay}, G. and {Steeghs}, D.},
        title = "{Optical evolution of AT 2024wpp: the high-velocity outflows in Cow-like transients are consistent with high spherical symmetry}",
      journal = {\mnras},
         year = 2025,
        month = mar,
       volume = {537},
       number = {4},
        pages = {3298-3309},
          doi = {10.1093/mnras/staf232},
archivePrefix = {arXiv},
       eprint = {2411.03272},
 primaryClass = {astro-ph.HE},
       adsurl = {https://ui.adsabs.harvard.edu/abs/2025MNRAS.537.3298P}
}

@ARTICLE{1999ApJ...524..262M,
       author = {{MacFadyen}, A.~I. and {Woosley}, S.~E.},
        title = "{Collapsars: Gamma-Ray Bursts and Explosions in ``Failed Supernovae''}",
      journal = {\apj},
         year = 1999,
        month = oct,
       volume = {524},
       number = {1},
        pages = {262-289},
          doi = {10.1086/307790},
archivePrefix = {arXiv},
       eprint = {astro-ph/9810274},
 primaryClass = {astro-ph},
       adsurl = {https://ui.adsabs.harvard.edu/abs/1999ApJ...524..262M}
}

@ARTICLE{2023MNRAS.520..879M,
       author = {{Metha}, Benjamin and {Trenti}, Michele},
        title = "{The internal metallicity distributions of simulated galaxies from EAGLE, Illustris, and IllustrisTNG at z = 1.8-4 as probed by gamma-ray burst hosts}",
      journal = {\mnras},
         year = 2023,
        month = mar,
       volume = {520},
       number = {1},
        pages = {879-896},
          doi = {10.1093/mnras/stad165},
archivePrefix = {arXiv},
       eprint = {2301.07131},
 primaryClass = {astro-ph.GA},
       adsurl = {https://ui.adsabs.harvard.edu/abs/2023MNRAS.520..879M}
}

@ARTICLE{2015A&A...582A..78M,
       author = {{Micha{\l}owski}, M.~J. and {Gentile}, G. and {Hjorth}, J. and {Krumholz}, M.~R. and {Tanvir}, N.~R. and {Kamphuis}, P. and {Burlon}, D. and {Baes}, M. and {Basa}, S. and {Berta}, S. and {Castro Cer{\'o}n}, J.~M. and {Crosby}, D. and {D'Elia}, V. and {Elliott}, J. and {Greiner}, J. and {Hunt}, L.~K. and {Klose}, S. and {Koprowski}, M.~P. and {Le Floc'h}, E. and {Malesani}, D. and {Murphy}, T. and {Nicuesa Guelbenzu}, A. and {Palazzi}, E. and {Rasmussen}, J. and {Rossi}, A. and {Savaglio}, S. and {Schady}, P. and {Sollerman}, J. and {de Ugarte Postigo}, A. and {Watson}, D. and {van der Werf}, P. and {Vergani}, S.~D. and {Xu}, D.},
        title = "{Massive stars formed in atomic hydrogen reservoirs: H I observations of gamma-ray burst host galaxies}",
      journal = {\aap},
         year = 2015,
        month = oct,
       volume = {582},
          eid = {A78},
        pages = {A78},
          doi = {10.1051/0004-6361/201526542},
archivePrefix = {arXiv},
       eprint = {1508.03094},
 primaryClass = {astro-ph.GA},
       adsurl = {https://ui.adsabs.harvard.edu/abs/2015A&A...582A..78M}
}

@ARTICLE{2020MNRAS.495..992L,
       author = {{Lyman}, J.~D. and {Galbany}, L. and {S{\'a}nchez}, S.~F. and {Anderson}, J.~P. and {Kuncarayakti}, H. and {Prieto}, J.~L.},
        title = "{Studying the environment of AT 2018cow with MUSE}",
      journal = {\mnras},
         year = 2020,
        month = jun,
       volume = {495},
       number = {1},
        pages = {992-999},
          doi = {10.1093/mnras/staa1243},
archivePrefix = {arXiv},
       eprint = {2005.02412},
 primaryClass = {astro-ph.GA},
       adsurl = {https://ui.adsabs.harvard.edu/abs/2020MNRAS.495..992L}
}

@ARTICLE{2020MNRAS.491.3479C,
       author = {{Chrimes}, A.~A. and {Stanway}, E.~R. and {Eldridge}, J.~J.},
        title = "{Binary population synthesis models for core-collapse gamma-ray burst progenitors}",
      journal = {\mnras},
         year = 2020,
        month = jan,
       volume = {491},
       number = {3},
        pages = {3479-3495},
          doi = {10.1093/mnras/stz3246},
archivePrefix = {arXiv},
       eprint = {1911.08387},
 primaryClass = {astro-ph.HE},
       adsurl = {https://ui.adsabs.harvard.edu/abs/2020MNRAS.491.3479C}
}

@ARTICLE{2020ApJ...895...49H,
       author = {{Ho}, Anna Y.~Q. and {Perley}, Daniel A. and {Kulkarni}, S.~R. and {Dong}, Dillon Z.~J. and {De}, Kishalay and {Chandra}, Poonam and {Andreoni}, Igor and {Bellm}, Eric C. and {Burdge}, Kevin B. and {Coughlin}, Michael and {Dekany}, Richard and {Feeney}, Michael and {Frederiks}, Dmitry D. and {Fremling}, Christoffer and {Golkhou}, V. Zach and {Graham}, Matthew J. and {Hale}, David and {Helou}, George and {Horesh}, Assaf and {Kasliwal}, Mansi M. and {Laher}, Russ R. and {Masci}, Frank J. and {Miller}, A.~A. and {Porter}, Michael and {Ridnaia}, Anna and {Rusholme}, Ben and {Shupe}, David L. and {Soumagnac}, Maayane T. and {Svinkin}, Dmitry S.},
        title = "{The Koala: A Fast Blue Optical Transient with Luminous Radio Emission from a Starburst Dwarf Galaxy at z = 0.27}",
      journal = {\apj},
         year = 2020,
        month = may,
       volume = {895},
       number = {1},
          eid = {49},
        pages = {49},
          doi = {10.3847/1538-4357/ab8bcf},
archivePrefix = {arXiv},
       eprint = {2003.01222},
 primaryClass = {astro-ph.HE},
       adsurl = {https://ui.adsabs.harvard.edu/abs/2020ApJ...895...49H}
}

@ARTICLE{2019ApJ...871...73H,
       author = {{Ho}, Anna Y.~Q. and {Phinney}, E. Sterl and {Ravi}, Vikram and {Kulkarni}, S.~R. and {Petitpas}, Glen and {Emonts}, Bjorn and {Bhalerao}, V. and {Blundell}, Ray and {Cenko}, S. Bradley and {Dobie}, Dougal and {Howie}, Ryan and {Kamraj}, Nikita and {Kasliwal}, Mansi M. and {Murphy}, Tara and {Perley}, Daniel A. and {Sridharan}, T.~K. and {Yoon}, Ilsang},
        title = "{AT2018cow: A Luminous Millimeter Transient}",
      journal = {\apj},
         year = 2019,
        month = jan,
       volume = {871},
       number = {1},
          eid = {73},
        pages = {73},
          doi = {10.3847/1538-4357/aaf473},
archivePrefix = {arXiv},
       eprint = {1810.10880},
 primaryClass = {astro-ph.HE},
       adsurl = {https://ui.adsabs.harvard.edu/abs/2019ApJ...871...73H}
}

@ARTICLE{2023MNRAS.521.3323M,
       author = {{Maund}, Justyn R. and {H{\"o}flich}, Peter A. and {Steele}, Iain A. and {Yang(杨轶)}, Yi and {Wiersema}, Klaas and {Kobayashi}, Shiho and {Jordana-Mitjans}, Nuria and {Mundell}, Carole and {Gomboc}, Andreja and {Guidorzi}, Cristiano and {Smith}, Robert J.},
        title = "{A flash of polarized optical light points to an aspherical 'cow'}",
      journal = {\mnras},
         year = 2023,
        month = may,
       volume = {521},
       number = {3},
        pages = {3323-3332},
          doi = {10.1093/mnras/stad539},
archivePrefix = {arXiv},
       eprint = {2303.00787},
 primaryClass = {astro-ph.SR},
       adsurl = {https://ui.adsabs.harvard.edu/abs/2023MNRAS.521.3323M}
}

@ARTICLE{2015MNRAS.451.2656K,
       author = {{Kashiyama}, Kazumi and {Quataert}, Eliot},
        title = "{Fast luminous blue transients from newborn black holes}",
      journal = {\mnras},
         year = 2015,
        month = aug,
       volume = {451},
       number = {3},
        pages = {2656-2662},
          doi = {10.1093/mnras/stv1164},
archivePrefix = {arXiv},
       eprint = {1504.05582},
 primaryClass = {astro-ph.HE},
       adsurl = {https://ui.adsabs.harvard.edu/abs/2015MNRAS.451.2656K}
}

@ARTICLE{2024ApJ...968...14R,
       author = {{Rastinejad}, J.~C. and {Fong}, W. and {Levan}, A.~J. and {Tanvir}, N.~R. and {Kilpatrick}, C.~D. and {Fruchter}, A.~S. and {Anand}, S. and {Bhirombhakdi}, K. and {Covino}, S. and {Fynbo}, J.~P.~U. and {Halevi}, G. and {Hartmann}, D.~H. and {Heintz}, K.~E. and {Izzo}, L. and {Jakobsson}, P. and {Kangas}, T. and {Lamb}, G.~P. and {Malesani}, D.~B. and {Melandri}, A. and {Metzger}, B.~D. and {Milvang-Jensen}, B. and {Pian}, E. and {Pugliese}, G. and {Rossi}, A. and {Siegel}, D.~M. and {Singh}, P. and {Stratta}, G.},
        title = "{A Hubble Space Telescope Search for r-Process Nucleosynthesis in Gamma-Ray Burst Supernovae}",
      journal = {\apj},
         year = 2024,
        month = jun,
       volume = {968},
       number = {1},
          eid = {14},
        pages = {14},
          doi = {10.3847/1538-4357/ad409c},
archivePrefix = {arXiv},
       eprint = {2312.04630},
 primaryClass = {astro-ph.HE},
       adsurl = {https://ui.adsabs.harvard.edu/abs/2024ApJ...968...14R}
}

@ARTICLE{2025arXiv251009745K,
       author = {{Klencki}, Jakub and {Metzger}, Brian D.},
        title = "{Luminous Fast Blue Optical Transients as ``Failed'' Gravitational Wave Sources: Helium Core$-$Black Hole Mergers Following Delayed Dynamical Instability}",
      journal = {arXiv e-prints},
         year = 2025,
        month = oct,
          eid = {arXiv:2510.09745},
        pages = {arXiv:2510.09745},
          doi = {10.48550/arXiv.2510.09745},
archivePrefix = {arXiv},
       eprint = {2510.09745},
 primaryClass = {astro-ph.HE},
       adsurl = {https://ui.adsabs.harvard.edu/abs/2025arXiv251009745K}
}

@ARTICLE{2022MNRAS.514.1188B,
       author = {{Byrne}, R.~A. and {Fraser}, M.},
        title = "{Nothing to see here: failed supernovae are faint or rare}",
      journal = {\mnras},
         year = 2022,
        month = jul,
       volume = {514},
       number = {1},
        pages = {1188-1205},
          doi = {10.1093/mnras/stac1308},
archivePrefix = {arXiv},
       eprint = {2201.12187},
 primaryClass = {astro-ph.SR},
       adsurl = {https://ui.adsabs.harvard.edu/abs/2022MNRAS.514.1188B}
}

@ARTICLE{2017MNRAS.468.4968A,
       author = {{Adams}, S.~M. and {Kochanek}, C.~S. and {Gerke}, J.~R. and {Stanek}, K.~Z. and {Dai}, X.},
        title = "{The search for failed supernovae with the Large Binocular Telescope: confirmation of a disappearing star}",
      journal = {\mnras},
         year = 2017,
        month = jul,
       volume = {468},
       number = {4},
        pages = {4968-4981},
          doi = {10.1093/mnras/stx816},
archivePrefix = {arXiv},
       eprint = {1609.01283},
 primaryClass = {astro-ph.SR},
       adsurl = {https://ui.adsabs.harvard.edu/abs/2017MNRAS.468.4968A}
}

@ARTICLE{2025ApJ...979..117B,
       author = {{Beasor}, Emma R. and {Smith}, Nathan and {Jencson}, Jacob E.},
        title = "{The Red Supergiant Progenitor Luminosity Problem}",
      journal = {\apj},
         year = 2025,
        month = feb,
       volume = {979},
       number = {2},
          eid = {117},
        pages = {117},
          doi = {10.3847/1538-4357/ad8f3f},
archivePrefix = {arXiv},
       eprint = {2410.14027},
 primaryClass = {astro-ph.SR},
       adsurl = {https://ui.adsabs.harvard.edu/abs/2025ApJ...979..117B}
}

@ARTICLE{2014ApJ...785...28K,
       author = {{Kochanek}, C.~S.},
        title = "{Failed Supernovae Explain the Compact Remnant Mass Function}",
      journal = {\apj},
         year = 2014,
        month = apr,
       volume = {785},
       number = {1},
          eid = {28},
        pages = {28},
          doi = {10.1088/0004-637X/785/1/28},
archivePrefix = {arXiv},
       eprint = {1308.0013},
 primaryClass = {astro-ph.HE},
       adsurl = {https://ui.adsabs.harvard.edu/abs/2014ApJ...785...28K}
}

@ARTICLE{2014ApJ...781..119P,
       author = {{Perna}, Rosalba and {Duffell}, Paul and {Cantiello}, Matteo and {MacFadyen}, Andrew I.},
        title = "{The Fate of Fallback Matter around Newly Born Compact Objects}",
      journal = {\apj},
         year = 2014,
        month = feb,
       volume = {781},
       number = {2},
          eid = {119},
        pages = {119},
          doi = {10.1088/0004-637X/781/2/119},
archivePrefix = {arXiv},
       eprint = {1312.4981},
 primaryClass = {astro-ph.HE},
       adsurl = {https://ui.adsabs.harvard.edu/abs/2014ApJ...781..119P}
}

@ARTICLE{2022ApJ...932...84M,
       author = {{Metzger}, Brian D.},
        title = "{Luminous Fast Blue Optical Transients and Type Ibn/Icn SNe from Wolf-Rayet/Black Hole Mergers}",
      journal = {\apj},
         year = 2022,
        month = jun,
       volume = {932},
       number = {2},
          eid = {84},
        pages = {84},
          doi = {10.3847/1538-4357/ac6d59},
archivePrefix = {arXiv},
       eprint = {2203.04331},
 primaryClass = {astro-ph.HE},
       adsurl = {https://ui.adsabs.harvard.edu/abs/2022ApJ...932...84M}
}

@ARTICLE{2023A&A...676A..31D,
       author = {{Disberg}, P. and {Nelemans}, G.},
        title = "{Failed supernovae as a natural explanation for the binary black hole mass distribution}",
      journal = {\aap},
         year = 2023,
        month = aug,
       volume = {676},
          eid = {A31},
        pages = {A31},
          doi = {10.1051/0004-6361/202245693},
archivePrefix = {arXiv},
       eprint = {2306.14332},
 primaryClass = {astro-ph.HE},
       adsurl = {https://ui.adsabs.harvard.edu/abs/2023A&A...676A..31D}
}

@ARTICLE{2025A&A...700A..20M,
       author = {{Maltsev}, K. and {Schneider}, F.~R.~N. and {Mandel}, I. and {M{\"u}ller}, B. and {Heger}, A. and {R{\"o}pke}, F.~K. and {Laplace}, E.},
        title = "{Explodability criteria for the neutrino-driven supernova mechanism}",
      journal = {\aap},
         year = 2025,
        month = aug,
       volume = {700},
          eid = {A20},
        pages = {A20},
          doi = {10.1051/0004-6361/202554931},
archivePrefix = {arXiv},
       eprint = {2503.23856},
 primaryClass = {astro-ph.SR},
       adsurl = {https://ui.adsabs.harvard.edu/abs/2025A&A...700A..20M}
}

@ARTICLE{2024AandA...691A.329C,
       author = {{Chrimes}, A.~A. and {Coppejans}, D.~L. and {Jonker}, P.~G. and {Levan}, A.~J. and {Groot}, P.~J. and {Mummery}, A. and {Stanway}, E.~R.},
        title = "{Multi-wavelength observations of the luminous fast blue optical transient AT 2023fhn: Up to {\ensuremath{\sim}}200 days post-explosion}",
      journal = {\aap},
         year = 2024,
        month = nov,
       volume = {691},
          eid = {A329},
        pages = {A329},
          doi = {10.1051/0004-6361/202451172},
archivePrefix = {arXiv},
       eprint = {2406.13821},
 primaryClass = {astro-ph.HE},
       adsurl = {https://ui.adsabs.harvard.edu/abs/2024A&A...691A.329C}
}

@ARTICLE{2021MNRAS.508.5138P,
       author = {{Perley}, Daniel A. and {Ho}, Anna Y.~Q. and {Yao}, Yuhan and {Fremling}, Christoffer and {Anderson}, Joseph P. and {Schulze}, Steve and {Kumar}, Harsh and {Anupama}, G.~C. and {Barway}, Sudhanshu and {Bellm}, Eric C. and {Bhalerao}, Varun and {Chen}, Ting-Wan and {Duev}, Dmitry A. and {Galbany}, Llu{\'\i}s and {Graham}, Matthew J. and {Gromadzki}, Mariusz and {Guti{\'e}rrez}, Claudia P. and {Ihanec}, Nada and {Inserra}, Cosimo and {Kasliwal}, Mansi M. and {Kool}, Erik C. and {Kulkarni}, S.~R. and {Laher}, Russ R. and {Masci}, Frank J. and {Neill}, James D. and {Nicholl}, Matt and {Pursiainen}, Miika and {van Roestel}, Joannes and {Sharma}, Yashvi and {Sollerman}, Jesper and {Walters}, Richard and {Wiseman}, Philip},
        title = "{Real-time discovery of AT2020xnd: a fast, luminous ultraviolet transient with minimal radioactive ejecta}",
      journal = {\mnras},
         year = 2021,
        month = dec,
       volume = {508},
       number = {4},
        pages = {5138-5147},
          doi = {10.1093/mnras/stab2785},
archivePrefix = {arXiv},
       eprint = {2103.01968},
 primaryClass = {astro-ph.HE},
       adsurl = {https://ui.adsabs.harvard.edu/abs/2021MNRAS.508.5138P}
}

@ARTICLE{2004ApJ...613..898T,
       author = {{Tremonti}, Christy A. and {Heckman}, Timothy M. and {Kauffmann}, Guinevere and {Brinchmann}, Jarle and {Charlot}, St{\'e}phane and {White}, Simon D.~M. and {Seibert}, Mark and {Peng}, Eric W. and {Schlegel}, David J. and {Uomoto}, Alan and {Fukugita}, Masataka and {Brinkmann}, Jon},
        title = "{The Origin of the Mass-Metallicity Relation: Insights from 53,000 Star-forming Galaxies in the Sloan Digital Sky Survey}",
      journal = {\apj},
         year = 2004,
        month = oct,
       volume = {613},
       number = {2},
        pages = {898-913},
          doi = {10.1086/423264},
archivePrefix = {arXiv},
       eprint = {astro-ph/0405537},
 primaryClass = {astro-ph},
       adsurl = {https://ui.adsabs.harvard.edu/abs/2004ApJ...613..898T}
}

@ARTICLE{2023ApJ...948..105V,
       author = {{van Son}, L.~A.~C. and {de Mink}, S.~E. and {Chru{\'s}li{\'n}ska}, M. and {Conroy}, C. and {Pakmor}, R. and {Hernquist}, L.},
        title = "{The Locations of Features in the Mass Distribution of Merging Binary Black Holes Are Robust against Uncertainties in the Metallicity-dependent Cosmic Star Formation History}",
      journal = {\apj},
         year = 2023,
        month = may,
       volume = {948},
       number = {2},
          eid = {105},
        pages = {105},
          doi = {10.3847/1538-4357/acbf51},
archivePrefix = {arXiv},
       eprint = {2209.03385},
 primaryClass = {astro-ph.GA},
       adsurl = {https://ui.adsabs.harvard.edu/abs/2023ApJ...948..105V}
}

@ARTICLE{2025ApJ...979..209V,
       author = {{van Son}, L.~A.~C. and {Roy}, S.~K. and {Mandel}, I. and {Farr}, W.~M. and {Lam}, A. and {Merritt}, J. and {Broekgaarden}, F.~S. and {Sander}, A.~A.~C. and {Andrews}, J.~J.},
        title = "{Not Just Winds: Why Models Find That Binary Black Hole Formation Is Metallicity-dependent, while Binary Neutron Star Formation Is Not}",
      journal = {\apj},
         year = 2025,
        month = feb,
       volume = {979},
       number = {2},
          eid = {209},
        pages = {209},
          doi = {10.3847/1538-4357/ada14a},
archivePrefix = {arXiv},
       eprint = {2411.02484},
 primaryClass = {astro-ph.HE},
       adsurl = {https://ui.adsabs.harvard.edu/abs/2025ApJ...979..209V}
}

@ARTICLE{2023RNAAS...7..126M,
       author = {{Matthews}, D. and {Margutti}, R. and {Metzger}, B.~D. and {Milisavljevic}, D. and {Migliori}, G. and {Laskar}, T. and {Brethauer}, D. and {Berger}, E. and {Chornock}, R. and {Drout}, M. and {Ramirez-Ruiz}, E.},
        title = "{Unprecedented X-Ray Emission from the Fast Blue Optical Transient AT2022tsd}",
      journal = {Research Notes of the American Astronomical Society},
         year = 2023,
        month = jun,
       volume = {7},
       number = {6},
          eid = {126},
        pages = {126},
          doi = {10.3847/2515-5172/acdde1},
archivePrefix = {arXiv},
       eprint = {2306.01114},
 primaryClass = {astro-ph.HE},
       adsurl = {https://ui.adsabs.harvard.edu/abs/2023RNAAS...7..126M}
}

@ARTICLE{2024ApJ...974...67L,
       author = {{Linial}, Itai and {Quataert}, Eliot},
        title = "{Tidal Disruption of a Star on a Nearly Circular Orbit}",
      journal = {\apj},
         year = 2024,
        month = oct,
       volume = {974},
       number = {1},
          eid = {67},
        pages = {67},
          doi = {10.3847/1538-4357/ad67cf},
archivePrefix = {arXiv},
       eprint = {2407.00149},
 primaryClass = {astro-ph.HE},
       adsurl = {https://ui.adsabs.harvard.edu/abs/2024ApJ...974...67L}
}

@ARTICLE{2022MNRAS.513.3810G,
       author = {{Gottlieb}, Ore and {Tchekhovskoy}, Alexander and {Margutti}, Raffaella},
        title = "{Shocked jets in CCSNe can power the zoo of fast blue optical transients}",
      journal = {\mnras},
         year = 2022,
        month = jul,
       volume = {513},
       number = {3},
        pages = {3810-3817},
          doi = {10.1093/mnras/stac910},
archivePrefix = {arXiv},
       eprint = {2201.04636},
 primaryClass = {astro-ph.HE},
       adsurl = {https://ui.adsabs.harvard.edu/abs/2022MNRAS.513.3810G}
}

@ARTICLE{2019ApJ...872...18M,
       author = {{Margutti}, R. and {Metzger}, B.~D. and {Chornock}, R. and {Vurm}, I. and {Roth}, N. and {Grefenstette}, B.~W. and {Savchenko}, V. and {Cartier}, R. and {Steiner}, J.~F. and {Terreran}, G. and {Margalit}, B. and {Migliori}, G. and {Milisavljevic}, D. and {Alexander}, K.~D. and {Bietenholz}, M. and {Blanchard}, P.~K. and {Bozzo}, E. and {Brethauer}, D. and {Chilingarian}, I.~V. and {Coppejans}, D.~L. and {Ducci}, L. and {Ferrigno}, C. and {Fong}, W. and {G{\"o}tz}, D. and {Guidorzi}, C. and {Hajela}, A. and {Hurley}, K. and {Kuulkers}, E. and {Laurent}, P. and {Mereghetti}, S. and {Nicholl}, M. and {Patnaude}, D. and {Ubertini}, P. and {Banovetz}, J. and {Bartel}, N. and {Berger}, E. and {Coughlin}, E.~R. and {Eftekhari}, T. and {Frederiks}, D.~D. and {Kozlova}, A.~V. and {Laskar}, T. and {Svinkin}, D.~S. and {Drout}, M.~R. and {MacFadyen}, A. and {Paterson}, K.},
        title = "{An Embedded X-Ray Source Shines through the Aspherical AT 2018cow: Revealing the Inner Workings of the Most Luminous Fast-evolving Optical Transients}",
      journal = {\apj},
         year = 2019,
        month = feb,
       volume = {872},
       number = {1},
          eid = {18},
        pages = {18},
          doi = {10.3847/1538-4357/aafa01},
archivePrefix = {arXiv},
       eprint = {1810.10720},
 primaryClass = {astro-ph.HE},
       adsurl = {https://ui.adsabs.harvard.edu/abs/2019ApJ...872...18M}
}

@ARTICLE{2018MNRAS.479...75S,
       author = {{Stanway}, E.~R. and {Eldridge}, J.~J.},
        title = "{Re-evaluating old stellar populations}",
      journal = {\mnras},
         year = 2018,
        month = sep,
       volume = {479},
       number = {1},
        pages = {75-93},
          doi = {10.1093/mnras/sty1353},
archivePrefix = {arXiv},
       eprint = {1805.08784},
 primaryClass = {astro-ph.GA},
       adsurl = {https://ui.adsabs.harvard.edu/abs/2018MNRAS.479...75S}
}

@ARTICLE{2020ApJ...890...51E,
       author = {{Ertl}, T. and {Woosley}, S.~E. and {Sukhbold}, Tuguldur and {Janka}, H. -T.},
        title = "{The Explosion of Helium Stars Evolved with Mass Loss}",
      journal = {\apj},
         year = 2020,
        month = feb,
       volume = {890},
       number = {1},
          eid = {51},
        pages = {51},
          doi = {10.3847/1538-4357/ab6458},
archivePrefix = {arXiv},
       eprint = {1910.01641},
 primaryClass = {astro-ph.HE},
       adsurl = {https://ui.adsabs.harvard.edu/abs/2020ApJ...890...51E}
}

@ARTICLE{2025A&A...695A..71L,
       author = {{Laplace}, E. and {Schneider}, F.~R.~N. and {Podsiadlowski}, Ph.},
        title = "{It's written in the massive stars: The role of stellar physics in the formation of black holes}",
      journal = {\aap},
         year = 2025,
        month = mar,
       volume = {695},
          eid = {A71},
        pages = {A71},
          doi = {10.1051/0004-6361/202451077},
archivePrefix = {arXiv},
       eprint = {2409.02058},
 primaryClass = {astro-ph.SR},
       adsurl = {https://ui.adsabs.harvard.edu/abs/2025A&A...695A..71L}
}

@ARTICLE{2009ARA&A..47...63S,
       author = {{Smartt}, Stephen J.},
        title = "{Progenitors of Core-Collapse Supernovae}",
      journal = {\araa},
         year = 2009,
        month = sep,
       volume = {47},
       number = {1},
        pages = {63-106},
          doi = {10.1146/annurev-astro-082708-101737},
archivePrefix = {arXiv},
       eprint = {0908.0700},
 primaryClass = {astro-ph.SR},
       adsurl = {https://ui.adsabs.harvard.edu/abs/2009ARA&A..47...63S}
}

@ARTICLE{2020MNRAS.499.2803P,
       author = {{Patton}, Rachel A. and {Sukhbold}, Tuguldur},
        title = "{Towards a realistic explosion landscape for binary population synthesis}",
      journal = {\mnras},
         year = 2020,
        month = dec,
       volume = {499},
       number = {2},
        pages = {2803-2816},
          doi = {10.1093/mnras/staa3029},
archivePrefix = {arXiv},
       eprint = {2005.03055},
 primaryClass = {astro-ph.SR},
       adsurl = {https://ui.adsabs.harvard.edu/abs/2020MNRAS.499.2803P}
}

@ARTICLE{2020MNRAS.492.2578S,
       author = {{Sukhbold}, Tuguldur and {Adams}, Scott},
        title = "{Missing red supergiants and carbon burning}",
      journal = {\mnras},
         year = 2020,
        month = feb,
       volume = {492},
       number = {2},
        pages = {2578-2587},
          doi = {10.1093/mnras/staa059},
archivePrefix = {arXiv},
       eprint = {1905.00474},
 primaryClass = {astro-ph.HE},
       adsurl = {https://ui.adsabs.harvard.edu/abs/2020MNRAS.492.2578S}
}

@ARTICLE{2001MNRAS.322..231K,
       author = {{Kroupa}, Pavel},
        title = "{On the variation of the initial mass function}",
      journal = {\mnras},
         year = 2001,
        month = apr,
       volume = {322},
       number = {2},
        pages = {231-246},
          doi = {10.1046/j.1365-8711.2001.04022.x},
archivePrefix = {arXiv},
       eprint = {astro-ph/0009005},
 primaryClass = {astro-ph},
       adsurl = {https://ui.adsabs.harvard.edu/abs/2001MNRAS.322..231K}
}

@ARTICLE{2017ApJS..230...15M,
       author = {{Moe}, Maxwell and {Di Stefano}, Rosanne},
        title = "{Mind Your Ps and Qs: The Interrelation between Period (P) and Mass-ratio (Q) Distributions of Binary Stars}",
      journal = {\apjs},
         year = 2017,
        month = jun,
       volume = {230},
       number = {2},
          eid = {15},
        pages = {15},
          doi = {10.3847/1538-4365/aa6fb6},
archivePrefix = {arXiv},
       eprint = {1606.05347},
 primaryClass = {astro-ph.SR},
       adsurl = {https://ui.adsabs.harvard.edu/abs/2017ApJS..230...15M}
}

@ARTICLE{2017PASA...34...58E,
       author = {{Eldridge}, J.~J. and {Stanway}, E.~R. and {Xiao}, L. and {McClelland}, L.~A.~S. and {Taylor}, G. and {Ng}, M. and {Greis}, S.~M.~L. and {Bray}, J.~C.},
        title = "{Binary Population and Spectral Synthesis Version 2.1: Construction, Observational Verification, and New Results}",
      journal = {\pasa},
         year = 2017,
        month = nov,
       volume = {34},
          eid = {e058},
        pages = {e058},
          doi = {10.1017/pasa.2017.51},
archivePrefix = {arXiv},
       eprint = {1710.02154},
 primaryClass = {astro-ph.SR},
       adsurl = {https://ui.adsabs.harvard.edu/abs/2017PASA...34...58E}
}

@ARTICLE{2022MNRAS.511..903P,
       author = {{Patton}, Rachel A. and {Sukhbold}, Tuguldur and {Eldridge}, J.~J.},
        title = "{Comparing compact object distributions from mass- and presupernova core structure-based prescriptions}",
      journal = {\mnras},
         year = 2022,
        month = mar,
       volume = {511},
       number = {1},
        pages = {903-913},
          doi = {10.1093/mnras/stab3797},
archivePrefix = {arXiv},
       eprint = {2106.05978},
 primaryClass = {astro-ph.HE},
       adsurl = {https://ui.adsabs.harvard.edu/abs/2022MNRAS.511..903P}
}

@ARTICLE{2024PhRvD.110h3024D,
       author = {{Dean}, Coleman and {Fern{\'a}ndez}, Rodrigo},
        title = "{Collapsar disk outflows: Heavy element production}",
      journal = {\prd},
         year = 2024,
        month = oct,
       volume = {110},
       number = {8},
          eid = {083024},
        pages = {083024},
          doi = {10.1103/PhysRevD.110.083024},
archivePrefix = {arXiv},
       eprint = {2408.15338},
 primaryClass = {astro-ph.HE},
       adsurl = {https://ui.adsabs.harvard.edu/abs/2024PhRvD.110h3024D}
}

@ARTICLE{2019MNRAS.485L..83Q,
       author = {{Quataert}, E. and {Lecoanet}, D. and {Coughlin}, E.~R.},
        title = "{Black hole accretion discs and luminous transients in failed supernovae from non-rotating supergiants}",
      journal = {\mnras},
         year = 2019,
        month = may,
       volume = {485},
       number = {1},
        pages = {L83-L88},
          doi = {10.1093/mnrasl/slz031},
archivePrefix = {arXiv},
       eprint = {1811.12427},
 primaryClass = {astro-ph.SR},
       adsurl = {https://ui.adsabs.harvard.edu/abs/2019MNRAS.485L..83Q}
}

@ARTICLE{2025arXiv250818083T,
       author = {{The LIGO Scientific Collaboration} and {the Virgo Collaboration} and {the KAGRA Collaboration}},
        title = "{GWTC-4.0: Population Properties of Merging Compact Binaries}",
      journal = {arXiv e-prints},
         year = 2025,
        month = aug,
          eid = {arXiv:2508.18083},
        pages = {arXiv:2508.18083},
          doi = {10.48550/arXiv.2508.18083},
archivePrefix = {arXiv},
       eprint = {2508.18083},
 primaryClass = {astro-ph.HE},
       adsurl = {https://ui.adsabs.harvard.edu/abs/2025arXiv250818083T}

}

@ARTICLE{2024ApJ...964..171B,
       author = {{Beasor}, Emma R. and {Hosseinzadeh}, Griffin and {Smith}, Nathan and {Davies}, Ben and {Jencson}, Jacob E. and {Pearson}, Jeniveve and {Sand}, David J.},
        title = "{JWST Reveals a Luminous Infrared Source at the Position of the Failed Supernova Candidate N6946-BH1}",
      journal = {\apj},
         year = 2024,
        month = apr,
       volume = {964},
       number = {2},
          eid = {171},
        pages = {171},
          doi = {10.3847/1538-4357/ad21fa},
archivePrefix = {arXiv},
       eprint = {2309.16121},
 primaryClass = {astro-ph.SR},
       adsurl = {https://ui.adsabs.harvard.edu/abs/2024ApJ...964..171B}
}

@ARTICLE{2018MNRAS.476.2366F,
       author = {{Fern{\'a}ndez}, Rodrigo and {Quataert}, Eliot and {Kashiyama}, Kazumi and {Coughlin}, Eric R.},
        title = "{Mass ejection in failed supernovae: variation with stellar progenitor}",
      journal = {\mnras},
         year = 2018,
        month = may,
       volume = {476},
       number = {2},
        pages = {2366-2383},
          doi = {10.1093/mnras/sty306},
archivePrefix = {arXiv},
       eprint = {1710.01735},
 primaryClass = {astro-ph.HE},
       adsurl = {https://ui.adsabs.harvard.edu/abs/2018MNRAS.476.2366F}
}

@ARTICLE{2025ApJ...986...84T,
       author = {{Tsuna}, Daichi and {Lu}, Wenbin},
        title = "{Stellar Tidal Disruptions by Newborn Neutron Stars or Black Holes: A Mechanism for Hydrogen-poor (Super)luminous Supernovae and Fast Blue Optical Transients}",
      journal = {\apj},
         year = 2025,
        month = jun,
       volume = {986},
       number = {1},
          eid = {84},
        pages = {84},
          doi = {10.3847/1538-4357/add158},
archivePrefix = {arXiv},
       eprint = {2501.03316},
 primaryClass = {astro-ph.HE},
       adsurl = {https://ui.adsabs.harvard.edu/abs/2025ApJ...986...84T}
}

@ARTICLE{2025ApJ...995..228S,
       author = {{Somalwar}, Jean J. and {Ravi}, Vikram and {Margutti}, Raffaella and {Chornock}, Ryan and {Natarajan}, Priyamvada and {Lu}, Wenbin and {Angus}, Charlotte and {Graham}, Matthew J. and {Hammerstein}, Erica and {Nathan}, Edward and {Nicholl}, Matt and {Sharma}, Kritti and {Stein}, Robert and {Verdi}, Frank and {Yao}, Yuhan and {Bellm}, Eric C. and {Chen}, Tracy X. and {Coughlin}, Michael W. and {Hale}, David and {Kasliwal}, Mansi M. and {Laher}, Russ R. and {Riddle}, Reed and {Sollerman}, Jesper},
        title = "{A Luminous and Hot Infrared through X-Ray Transient at a 5 kpc Offset from a Dwarf Galaxy}",
      journal = {\apj},
         year = 2025,
        month = dec,
       volume = {995},
       number = {2},
          eid = {228},
        pages = {228},
          doi = {10.3847/1538-4357/ae1501},
archivePrefix = {arXiv},
       eprint = {2505.11597},
 primaryClass = {astro-ph.HE},
       adsurl = {https://ui.adsabs.harvard.edu/abs/2025ApJ...995..228S}
}

@ARTICLE{2025A&A...695A..86N,
       author = {{Nersesian}, Angelos and {van der Wel}, Arjen and {Gallazzi}, Anna R. and {Kaushal}, Yasha and {Bezanson}, Rachel and {Zibetti}, Stefano and {Bell}, Eric F. and {D'Eugenio}, Francesco and {Leja}, Joel and {Martorano}, Marco and {Wu}, Po-Feng},
        title = "{More is better: Strong constraints on the stellar properties of LEGA-C z {\ensuremath{\sim}} 1 galaxies with Prospector}",
      journal = {\aap},
         year = 2025,
        month = mar,
       volume = {695},
          eid = {A86},
        pages = {A86},
          doi = {10.1051/0004-6361/202452662},
archivePrefix = {arXiv},
       eprint = {2502.03021},
 primaryClass = {astro-ph.GA},
       adsurl = {https://ui.adsabs.harvard.edu/abs/2025A&A...695A..86N}
}

@ARTICLE{2009MNRAS.395.1409S,
       author = {{Smartt}, S.~J. and {Eldridge}, J.~J. and {Crockett}, R.~M. and {Maund}, J.~R.},
        title = "{The death of massive stars - I. Observational constraints on the progenitors of Type II-P supernovae}",
      journal = {\mnras},
         year = 2009,
        month = may,
       volume = {395},
       number = {3},
        pages = {1409-1437},
          doi = {10.1111/j.1365-2966.2009.14506.x},
archivePrefix = {arXiv},
       eprint = {0809.0403},
 primaryClass = {astro-ph},
       adsurl = {https://ui.adsabs.harvard.edu/abs/2009MNRAS.395.1409S}
}

@ARTICLE{2019MNRAS.488.5300C,
       author = {{Chruslinska}, Martyna and {Nelemans}, Gijs},
        title = "{Metallicity of stars formed throughout the cosmic history based on the observational properties of star-forming galaxies}",
      journal = {\mnras},
         year = 2019,
        month = oct,
       volume = {488},
       number = {4},
        pages = {5300-5326},
          doi = {10.1093/mnras/stz2057},
archivePrefix = {arXiv},
       eprint = {1907.11243},
 primaryClass = {astro-ph.GA},
       adsurl = {https://ui.adsabs.harvard.edu/abs/2019MNRAS.488.5300C}
}

@ARTICLE{2006ApJ...638L..63L,
       author = {{Langer}, N. and {Norman}, C.~A.},
        title = "{On the Collapsar Model of Long Gamma-Ray Bursts:Constraints from Cosmic Metallicity Evolution}",
      journal = {\apjl},
         year = 2006,
        month = feb,
       volume = {638},
       number = {2},
        pages = {L63-L66},
          doi = {10.1086/500363},
archivePrefix = {arXiv},
       eprint = {astro-ph/0512271},
 primaryClass = {astro-ph},
       adsurl = {https://ui.adsabs.harvard.edu/abs/2006ApJ...638L..63L}
}

@ARTICLE{2022MNRAS.512L..66S,
       author = {{Sun}, Ning-Chen and {Maund}, Justyn R. and {Crowther}, Paul A. and {Liu}, Liang-Duan},
        title = "{A hot and luminous source at the site of the fast transient AT2018cow at 2-3 yr after its explosion}",
      journal = {\mnras},
         year = 2022,
        month = may,
       volume = {512},
       number = {1},
        pages = {L66-L70},
          doi = {10.1093/mnrasl/slac023},
archivePrefix = {arXiv},
       eprint = {2203.01960},
 primaryClass = {astro-ph.HE},
       adsurl = {https://ui.adsabs.harvard.edu/abs/2022MNRAS.512L..66S}
}

@ARTICLE{2022MNRAS.515.2591C,
       author = {{Chrimes}, A.~A. and {Gompertz}, B.~P. and {Kann}, D.~A. and {van Marle}, A.~J. and {Eldridge}, J.~J. and {Groot}, P.~J. and {Laskar}, T. and {Levan}, A.~J. and {Nicholl}, M. and {Stanway}, E.~R. and {Wiersema}, K.},
        title = "{Towards an understanding of long gamma-ray burst environments through circumstellar medium population synthesis predictions}",
      journal = {\mnras},
         year = 2022,
        month = sep,
       volume = {515},
       number = {2},
        pages = {2591-2611},
          doi = {10.1093/mnras/stac1796},
archivePrefix = {arXiv},
       eprint = {2206.13595},
 primaryClass = {astro-ph.HE},
       adsurl = {https://ui.adsabs.harvard.edu/abs/2022MNRAS.515.2591C}
}

@ARTICLE{2023MNRAS.519.3785S,
       author = {{Sun}, Ning-Chen and {Maund}, Justyn R. and {Shao}, Yali and {Janiak}, Ida A.},
        title = "{An environmental analysis of the fast transient AT2018cow and implications for its progenitor and late-time brightness}",
      journal = {\mnras},
         year = 2023,
        month = mar,
       volume = {519},
       number = {3},
        pages = {3785-3797},
          doi = {10.1093/mnras/stac3773},
archivePrefix = {arXiv},
       eprint = {2210.01144},
 primaryClass = {astro-ph.GA},
       adsurl = {https://ui.adsabs.harvard.edu/abs/2023MNRAS.519.3785S}
}

@ARTICLE{2023MNRAS.525.4042I,
       author = {{Inkenhaag}, Anne and {Jonker}, Peter G. and {Levan}, Andrew J. and {Chrimes}, Ashley A. and {Mummery}, Andrew and {Perley}, Daniel A. and {Tanvir}, Nial R.},
        title = "{Late-time HST UV and optical observations of AT 2018cow: extracting a cow from its background}",
      journal = {\mnras},
         year = 2023,
        month = nov,
       volume = {525},
       number = {3},
        pages = {4042-4056},
          doi = {10.1093/mnras/stad2531},
archivePrefix = {arXiv},
       eprint = {2308.07381},
 primaryClass = {astro-ph.HE},
       adsurl = {https://ui.adsabs.harvard.edu/abs/2023MNRAS.525.4042I}
}

@ARTICLE{2014ARA&A..52..487S,
       author = {{Smith}, Nathan},
        title = "{Mass Loss: Its Effect on the Evolution and Fate of High-Mass Stars}",
      journal = {\araa},
         year = 2014,
        month = aug,
       volume = {52},
        pages = {487-528},
          doi = {10.1146/annurev-astro-081913-040025},
archivePrefix = {arXiv},
       eprint = {1402.1237},
 primaryClass = {astro-ph.SR},
       adsurl = {https://ui.adsabs.harvard.edu/abs/2014ARA&A..52..487S}
}

@ARTICLE{2023MNRAS.526..152K,
       author = {{Kuroda}, Takami and {Shibata}, Masaru},
        title = "{Failed supernova simulations beyond black hole formation}",
      journal = {\mnras},
         year = 2023,
        month = nov,
       volume = {526},
       number = {1},
        pages = {152-159},
          doi = {10.1093/mnras/stad2710},
archivePrefix = {arXiv},
       eprint = {2307.06192},
 primaryClass = {astro-ph.HE},
       adsurl = {https://ui.adsabs.harvard.edu/abs/2023MNRAS.526..152K}
}

@ARTICLE{2025MNRAS.544L.108I,
       author = {{Inkenhaag}, Anne and {Levan}, Andrew J. and {Mummery}, Andrew and {Jonker}, Peter G.},
        title = "{AT 2018cow at {\ensuremath{\sim}}5 years: additional evidence for a tidal disruption origin}",
      journal = {\mnras},
         year = 2025,
        month = nov,
       volume = {544},
       number = {1},
        pages = {L108-L112},
          doi = {10.1093/mnrasl/slaf107},
archivePrefix = {arXiv},
       eprint = {2510.08505},
 primaryClass = {astro-ph.HE},
       adsurl = {https://ui.adsabs.harvard.edu/abs/2025MNRAS.544L.108I}
}

@ARTICLE{2025MNRAS.544.2225M,
       author = {{Mummery}, Andrew and {Nathan}, Edward and {Ingram}, Adam and {Gardner}, M.},
        title = "{Fitting transients with discs (FITTED): a public light curve and spectral fitting package based on evolving relativistic discs}",
      journal = {\mnras},
         year = 2025,
        month = dec,
       volume = {544},
       number = {2},
        pages = {2225-2240},
          doi = {10.1093/mnras/staf1565},
archivePrefix = {arXiv},
       eprint = {2408.15048},
 primaryClass = {astro-ph.HE},
       adsurl = {https://ui.adsabs.harvard.edu/abs/2025MNRAS.544.2225M}
}

@ARTICLE{2024A&A...691A.228C,
       author = {{Cao}, Zheng and {Jonker}, Peter G. and {Wen}, Sixiang and {Zabludoff}, Ann I.},
        title = "{Slim-disk modeling reveals an accreting intermediate-mass black hole in the luminous fast blue optical transient AT2018cow}",
      journal = {\aap},
         year = 2024,
        month = nov,
       volume = {691},
          eid = {A228},
        pages = {A228},
          doi = {10.1051/0004-6361/202451297},
archivePrefix = {arXiv},
       eprint = {2409.17695},
 primaryClass = {astro-ph.HE},
       adsurl = {https://ui.adsabs.harvard.edu/abs/2024A&A...691A.228C}
}

@ARTICLE{2022NatAs...6..249P,
       author = {{Pasham}, Dheeraj R. and {Ho}, Wynn C.~G. and {Alston}, William and {Remillard}, Ronald and {Ng}, Mason and {Gendreau}, Keith and {Metzger}, Brian D. and {Altamirano}, Diego and {Chakrabarty}, Deepto and {Fabian}, Andrew and {Miller}, Jon and {Bult}, Peter and {Arzoumanian}, Zaven and {Steiner}, James F. and {Strohmayer}, Tod and {Tombesi}, Francesco and {Homan}, Jeroen and {Cackett}, Edward M. and {Harding}, Alice},
        title = "{Evidence for a compact object in the aftermath of the extragalactic transient AT2018cow}",
      journal = {Nature Astronomy},
         year = 2021,
        month = dec,
       volume = {6},
        pages = {249-258},
          doi = {10.1038/s41550-021-01524-8},
archivePrefix = {arXiv},
       eprint = {2112.04531},
 primaryClass = {astro-ph.HE},
       adsurl = {https://ui.adsabs.harvard.edu/abs/2022NatAs...6..249P}
}

@ARTICLE{2024ApJ...963L..24M,
       author = {{Migliori}, Giulia and {Margutti}, R. and {Metzger}, B.~D. and {Chornock}, R. and {Vignali}, C. and {Brethauer}, D. and {Coppejans}, D.~L. and {Maccarone}, T. and {Rivera Sandoval}, L. and {Bright}, J.~S. and {Laskar}, T. and {Milisavljevic}, D. and {Berger}, E. and {Nayana}, A.~J.},
        title = "{Roaring to Softly Whispering: X-Ray Emission after {\ensuremath{\sim}}3.7 yr at the Location of the Transient AT2018cow and Implications for Accretion-powered Scenarios}",
      journal = {\apjl},
         year = 2024,
        month = mar,
       volume = {963},
       number = {1},
          eid = {L24},
        pages = {L24},
          doi = {10.3847/2041-8213/ad2764},
       adsurl = {https://ui.adsabs.harvard.edu/abs/2024ApJ...963L..24M}
}

@ARTICLE{2003ApJ...591..288H,
       author = {{Heger}, A. and {Fryer}, C.~L. and {Woosley}, S.~E. and {Langer}, N. and {Hartmann}, D.~H.},
        title = "{How Massive Single Stars End Their Life}",
      journal = {\apj},
         year = 2003,
        month = jul,
       volume = {591},
       number = {1},
        pages = {288-300},
          doi = {10.1086/375341},
archivePrefix = {arXiv},
       eprint = {astro-ph/0212469},
 primaryClass = {astro-ph},
       adsurl = {https://ui.adsabs.harvard.edu/abs/2003ApJ...591..288H}
}

@ARTICLE{2013A&A...560A..29R,
       author = {{Ram{\'\i}rez-Agudelo}, O.~H. and {Sim{\'o}n-D{\'\i}az}, S. and {Sana}, H. and {de Koter}, A. and {Sab{\'\i}n-Sanjul{\'\i}an}, C. and {de Mink}, S.~E. and {Dufton}, P.~L. and {Gr{\"a}fener}, G. and {Evans}, C.~J. and {Herrero}, A. and {Langer}, N. and {Lennon}, D.~J. and {Ma{\'\i}z Apell{\'a}niz}, J. and {Markova}, N. and {Najarro}, F. and {Puls}, J. and {Taylor}, W.~D. and {Vink}, J.~S.},
        title = "{The VLT-FLAMES Tarantula Survey. XII. Rotational velocities of the single O-type stars}",
      journal = {\aap},
         year = 2013,
        month = dec,
       volume = {560},
          eid = {A29},
        pages = {A29},
          doi = {10.1051/0004-6361/201321986},
archivePrefix = {arXiv},
       eprint = {1309.2929},
 primaryClass = {astro-ph.SR},
       adsurl = {https://ui.adsabs.harvard.edu/abs/2013A&A...560A..29R}
}

@ARTICLE{2012A&A...540A..56P,
       author = {{Pilkington}, K. and {Few}, C.~G. and {Gibson}, B.~K. and {Calura}, F. and {Michel-Dansac}, L. and {Thacker}, R.~J. and {Moll{\'a}}, M. and {Matteucci}, F. and {Rahimi}, A. and {Kawata}, D. and {Kobayashi}, C. and {Brook}, C.~B. and {Stinson}, G.~S. and {Couchman}, H.~M.~P. and {Bailin}, J. and {Wadsley}, J.},
        title = "{Metallicity gradients in disks. Do galaxies form inside-out?}",
      journal = {\aap},
         year = 2012,
        month = apr,
       volume = {540},
          eid = {A56},
        pages = {A56},
          doi = {10.1051/0004-6361/201117466},
archivePrefix = {arXiv},
       eprint = {1201.6359},
 primaryClass = {astro-ph.GA},
       adsurl = {https://ui.adsabs.harvard.edu/abs/2012A&A...540A..56P}
}

@ARTICLE{2025ApJ...978L..39J,
       author = {{Ju}, Mengting and {Wang}, Xin and {Jones}, Tucker and {Bari{\v{s}}i{\'c}}, Ivana and {Nanayakkara}, Themiya and {Bundy}, Kevin and {Faucher-Gigu{\`e}re}, Claude-Andr{\'e} and {Feng}, Shuai and {Glazebrook}, Karl and {Henry}, Alaina and {Malkan}, Matthew A. and {Obreschkow}, Danail and {Roy}, Namrata and {Sanders}, Ryan L. and {Sun}, Xunda and {Treu}, Tommaso and {Zhou}, Qianqiao},
        title = "{MSA-3D: Metallicity Gradients in Galaxies at z {\ensuremath{\sim}} 1 with JWST/NIRSpec Slit-stepping Spectroscopy}",
      journal = {\apjl},
         year = 2025,
        month = jan,
       volume = {978},
       number = {2},
          eid = {L39},
        pages = {L39},
          doi = {10.3847/2041-8213/ada150},
archivePrefix = {arXiv},
       eprint = {2409.01616},
 primaryClass = {astro-ph.GA},
       adsurl = {https://ui.adsabs.harvard.edu/abs/2025ApJ...978L..39J}
}

@ARTICLE{2015A&A...580A..92R,
       author = {{Ram{\'\i}rez-Agudelo}, O.~H. and {Sana}, H. and {de Mink}, S.~E. and {H{\'e}nault-Brunet}, V. and {de Koter}, A. and {Langer}, N. and {Tramper}, F. and {Gr{\"a}fener}, G. and {Evans}, C.~J. and {Vink}, J.~S. and {Dufton}, P.~L. and {Taylor}, W.~D.},
        title = "{The VLT-FLAMES Tarantula Survey. XXI. Stellar spin rates of O-type spectroscopic binaries}",
      journal = {\aap},
         year = 2015,
        month = aug,
       volume = {580},
          eid = {A92},
        pages = {A92},
          doi = {10.1051/0004-6361/201425424},
archivePrefix = {arXiv},
       eprint = {1507.02286},
 primaryClass = {astro-ph.SR},
       adsurl = {https://ui.adsabs.harvard.edu/abs/2015A&A...580A..92R}
}

@ARTICLE{2023ApJ...955...43C,
       author = {{Chen}, Yuyang and {Drout}, Maria R. and {Piro}, Anthony L. and {Kilpatrick}, Charles D. and {Foley}, Ryan J. and {Rojas-Bravo}, C{\'e}sar and {Magee}, M.~R.},
        title = "{Late-time Hubble Space Telescope Observations of AT 2018cow. II. Evolution of a UV-bright Underlying Source 2-4 Yr Post-discovery}",
      journal = {\apj},
         year = 2023,
        month = sep,
       volume = {955},
       number = {1},
          eid = {43},
        pages = {43},
          doi = {10.3847/1538-4357/ace964},
archivePrefix = {arXiv},
       eprint = {2303.03501},
 primaryClass = {astro-ph.HE},
       adsurl = {https://ui.adsabs.harvard.edu/abs/2023ApJ...955...43C}
}

@ARTICLE{2024MNRAS.527L..47C,
       author = {{Chrimes}, A.~A. and {Jonker}, P.~G. and {Levan}, A.~J. and {Coppejans}, D.~L. and {Gaspari}, N. and {Gompertz}, B.~P. and {Groot}, P.~J. and {Malesani}, D.~B. and {Mummery}, A. and {Stanway}, E.~R. and {Wiersema}, K.},
        title = "{AT2023fhn (the Finch): a luminous fast blue optical transient at a large offset from its host galaxy}",
      journal = {\mnras},
         year = 2024,
        month = jan,
       volume = {527},
       number = {1},
        pages = {L47-L53},
          doi = {10.1093/mnrasl/slad145},
archivePrefix = {arXiv},
       eprint = {2307.01771},
 primaryClass = {astro-ph.HE},
       adsurl = {https://ui.adsabs.harvard.edu/abs/2024MNRAS.527L..47C}
}

@ARTICLE{2025ApJ...987..164B,
       author = {{Burrows}, Adam and {Wang}, Tianshu and {Vartanyan}, David},
        title = "{Channels of Stellar-mass Black Hole Formation}",
      journal = {\apj},
         year = 2025,
        month = jul,
       volume = {987},
       number = {2},
          eid = {164},
        pages = {164},
          doi = {10.3847/1538-4357/addd04},
archivePrefix = {arXiv},
       eprint = {2412.07831},
 primaryClass = {astro-ph.SR},
       adsurl = {https://ui.adsabs.harvard.edu/abs/2025ApJ...987..164B}
}

@ARTICLE{2019MNRAS.487.2505K,
       author = {{Kuin}, N. Paul M. and {Wu}, Kinwah and {Oates}, Samantha and {Lien}, Amy and {Emery}, Sam and {Kennea}, Jamie A. and {de Pasquale}, Massimiliano and {Han}, Qin and {Brown}, Peter J. and {Tohuvavohu}, Aaron and {Breeveld}, Alice and {Burrows}, David N. and {Cenko}, S. Bradley and {Campana}, Sergio and {Levan}, Andrew and {Markwardt}, Craig and {Osborne}, Julian P. and {Page}, Mat J. and {Page}, Kim L. and {Sbarufatti}, Boris and {Siegel}, Michael and {Troja}, Eleonora},
        title = "{Swift spectra of AT2018cow: a white dwarf tidal disruption event?}",
      journal = {\mnras},
         year = 2019,
        month = aug,
       volume = {487},
       number = {2},
        pages = {2505-2521},
          doi = {10.1093/mnras/stz053},
archivePrefix = {arXiv},
       eprint = {1808.08492},
 primaryClass = {astro-ph.HE},
       adsurl = {https://ui.adsabs.harvard.edu/abs/2019MNRAS.487.2505K}
}

@ARTICLE{2025ApJ...982..144N,
       author = {{Nugent}, Anya E. and {Ji}, Alexander P. and {Fong}, Wen-fai and {Shah}, Hilay and {van de Voort}, Freeke},
        title = "{Where Has All the R-process Gone? Timescales for Gamma-Ray Burst Kilonovae to Enrich Their Host Galaxies}",
      journal = {\apj},
         year = 2025,
        month = apr,
       volume = {982},
       number = {2},
          eid = {144},
        pages = {144},
          doi = {10.3847/1538-4357/adbb6a},
archivePrefix = {arXiv},
       eprint = {2410.00095},
 primaryClass = {astro-ph.HE},
       adsurl = {https://ui.adsabs.harvard.edu/abs/2025ApJ...982..144N}
}

@ARTICLE{2007CSE.....9c..21P,
       author = {{Perez}, Fernando and {Granger}, Brian E.},
        title = "{IPython: A System for Interactive Scientific Computing}",
      journal = {Computing in Science and Engineering},
         year = 2007,
        month = jan,
       volume = {9},
       number = {3},
        pages = {21-29},
          doi = {10.1109/MCSE.2007.53},
       adsurl = {https://ui.adsabs.harvard.edu/abs/2007CSE.....9c..21P}
}

@article{Waskom2021,
    doi = {10.21105/joss.03021},
    url = {https://doi.org/10.21105/joss.03021},
    year = {2021},
    publisher = {The Open Journal},
    volume = {6},
    number = {60},
    pages = {3021},
    author = {Michael L. Waskom},
    title = {seaborn: statistical data visualization},
    journal = {Journal of Open Source Software}
 }

@article{astropy:2013,
Adsurl = {http://adsabs.harvard.edu/abs/2013A%26A...558A..33A},
Archiveprefix = {arXiv},
Author = {{Astropy Collaboration} and {Robitaille}, T.~P. and {Tollerud}, E.~J. and {Greenfield}, P. and {Droettboom}, M. and {Bray}, E. and {Aldcroft}, T. and {Davis}, M. and {Ginsburg}, A. and {Price-Whelan}, A.~M. and {Kerzendorf}, W.~E. and {Conley}, A. and {Crighton}, N. and {Barbary}, K. and {Muna}, D. and {Ferguson}, H. and {Grollier}, F. and {Parikh}, M.~M. and {Nair}, P.~H. and {Unther}, H.~M. and {Deil}, C. and {Woillez}, J. and {Conseil}, S. and {Kramer}, R. and {Turner}, J.~E.~H. and {Singer}, L. and {Fox}, R. and {Weaver}, B.~A. and {Zabalza}, V. and {Edwards}, Z.~I. and {Azalee Bostroem}, K. and {Burke}, D.~J. and {Casey}, A.~R. and {Crawford}, S.~M. and {Dencheva}, N. and {Ely}, J. and {Jenness}, T. and {Labrie}, K. and {Lim}, P.~L. and {Pierfederici}, F. and {Pontzen}, A. and {Ptak}, A. and {Refsdal}, B. and {Servillat}, M. and {Streicher}, O.},
Doi = {10.1051/0004-6361/201322068},
Eid = {A33},
Eprint = {1307.6212},
Journal = {\aap},
Month = oct,
Pages = {A33},
Primaryclass = {astro-ph.IM},
Title = {{Astropy: A community Python package for astronomy}},
Volume = 558,
Year = 2013}

@article{astropy:2018,
Adsurl = {https://ui.adsabs.harvard.edu/#abs/2018AJ....156..123T},
Author = {{Price-Whelan}, A.~M. and {Sip{\H{o}}cz}, B.~M. and {G{\"u}nther}, H.~M. and {Lim}, P.~L. and {Crawford}, S.~M. and {Conseil}, S. and {Shupe}, D.~L. and {Craig}, M.~W. and {Dencheva}, N. and {Ginsburg}, A. and {VanderPlas}, J.~T. and {Bradley}, L.~D. and {P{\'e}rez-Su{\'a}rez}, D. and {de Val-Borro}, M. and {Paper Contributors}, (Primary and {Aldcroft}, T.~L. and {Cruz}, K.~L. and {Robitaille}, T.~P. and {Tollerud}, E.~J. and {Coordination Committee}, (Astropy and {Ardelean}, C. and {Babej}, T. and {Bach}, Y.~P. and {Bachetti}, M. and {Bakanov}, A.~V. and {Bamford}, S.~P. and {Barentsen}, G. and {Barmby}, P. and {Baumbach}, A. and {Berry}, K.~L. and {Biscani}, F. and {Boquien}, M. and {Bostroem}, K.~A. and {Bouma}, L.~G. and {Brammer}, G.~B. and {Bray}, E.~M. and {Breytenbach}, H. and {Buddelmeijer}, H. and {Burke}, D.~J. and {Calderone}, G. and {Cano Rodr{\'\i}guez}, J.~L. and {Cara}, M. and {Cardoso}, J.~V.~M. and {Cheedella}, S. and {Copin}, Y. and {Corrales}, L. and {Crichton}, D. and {D{\textquoteright}Avella}, D. and {Deil}, C. and {Depagne}, {\'E}. and {Dietrich}, J.~P. and {Donath}, A. and {Droettboom}, M. and {Earl}, N. and {Erben}, T. and {Fabbro}, S. and {Ferreira}, L.~A. and {Finethy}, T. and {Fox}, R.~T. and {Garrison}, L.~H. and {Gibbons}, S.~L.~J. and {Goldstein}, D.~A. and {Gommers}, R. and {Greco}, J.~P. and {Greenfield}, P. and {Groener}, A.~M. and {Grollier}, F. and {Hagen}, A. and {Hirst}, P. and {Homeier}, D. and {Horton}, A.~J. and {Hosseinzadeh}, G. and {Hu}, L. and {Hunkeler}, J.~S. and {Ivezi{\'c}}, {\v{Z}}. and {Jain}, A. and {Jenness}, T. and {Kanarek}, G. and {Kendrew}, S. and {Kern}, N.~S. and {Kerzendorf}, W.~E. and {Khvalko}, A. and {King}, J. and {Kirkby}, D. and {Kulkarni}, A.~M. and {Kumar}, A. and {Lee}, A. and {Lenz}, D. and {Littlefair}, S.~P. and {Ma}, Z. and {Macleod}, D.~M. and {Mastropietro}, M. and {McCully}, C. and {Montagnac}, S. and {Morris}, B.~M. and {Mueller}, M. and {Mumford}, S.~J. and {Muna}, D. and {Murphy}, N.~A. and {Nelson}, S. and {Nguyen}, G.~H. and {Ninan}, J.~P. and {N{\"o}the}, M. and {Ogaz}, S. and {Oh}, S. and {Parejko}, J.~K. and {Parley}, N. and {Pascual}, S. and {Patil}, R. and {Patil}, A.~A. and {Plunkett}, A.~L. and {Prochaska}, J.~X. and {Rastogi}, T. and {Reddy Janga}, V. and {Sabater}, J. and {Sakurikar}, P. and {Seifert}, M. and {Sherbert}, L.~E. and {Sherwood-Taylor}, H. and {Shih}, A.~Y. and {Sick}, J. and {Silbiger}, M.~T. and {Singanamalla}, S. and {Singer}, L.~P. and {Sladen}, P.~H. and {Sooley}, K.~A. and {Sornarajah}, S. and {Streicher}, O. and {Teuben}, P. and {Thomas}, S.~W. and {Tremblay}, G.~R. and {Turner}, J.~E.~H. and {Terr{\'o}n}, V. and {van Kerkwijk}, M.~H. and {de la Vega}, A. and {Watkins}, L.~L. and {Weaver}, B.~A. and {Whitmore}, J.~B. and {Woillez}, J. and {Zabalza}, V. and {Contributors}, (Astropy},
Doi = {10.3847/1538-3881/aabc4f},
Eid = {123},
Journal = {\aj},
Month = Sep,
Pages = {123},
Primaryclass = {astro-ph.IM},
Title = {{The Astropy Project: Building an Open-science Project and Status of the v2.0 Core Package}},
Volume = {156},
Year = 2018}

@ARTICLE{2020Natur.585..357H,
       author = {{Harris}, Charles R. and {Millman}, K. Jarrod and {van der Walt}, St{\'e}fan J. and {Gommers}, Ralf and {Virtanen}, Pauli and {Cournapeau}, David and {Wieser}, Eric and {Taylor}, Julian and {Berg}, Sebastian and {Smith}, Nathaniel J. and {Kern}, Robert and {Picus}, Matti and {Hoyer}, Stephan and {van Kerkwijk}, Marten H. and {Brett}, Matthew and {Haldane}, Allan and {del R{\'\i}o}, Jaime Fern{\'a}ndez and {Wiebe}, Mark and {Peterson}, Pearu and {G{\'e}rard-Marchant}, Pierre and {Sheppard}, Kevin and {Reddy}, Tyler and {Weckesser}, Warren and {Abbasi}, Hameer and {Gohlke}, Christoph and {Oliphant}, Travis E.},
        title = "{Array programming with NumPy}",
      journal = {\nat},
         year = 2020,
        month = sep,
       volume = {585},
       number = {7825},
        pages = {357-362},
          doi = {10.1038/s41586-020-2649-2},
archivePrefix = {arXiv},
       eprint = {2006.10256},
 primaryClass = {cs.MS},
       adsurl = {https://ui.adsabs.harvard.edu/abs/2020Natur.585..357H}
}

@ARTICLE{2020NatMe..17..261V,
       author = {{Virtanen}, Pauli and {Gommers}, Ralf and {Oliphant}, Travis E. and {Haberland}, Matt and {Reddy}, Tyler and {Cournapeau}, David and {Burovski}, Evgeni and {Peterson}, Pearu and {Weckesser}, Warren and {Bright}, Jonathan and {van der Walt}, St{\'e}fan J. and {Brett}, Matthew and {Wilson}, Joshua and {Millman}, K. Jarrod and {Mayorov}, Nikolay and {Nelson}, Andrew R.~J. and {Jones}, Eric and {Kern}, Robert and {Larson}, Eric and {Carey}, C.~J. and {Polat}, {\.I}lhan and {Feng}, Yu and {Moore}, Eric W. and {VanderPlas}, Jake and {Laxalde}, Denis and {Perktold}, Josef and {Cimrman}, Robert and {Henriksen}, Ian and {Quintero}, E.~A. and {Harris}, Charles R. and {Archibald}, Anne M. and {Ribeiro}, Ant{\^o}nio H. and {Pedregosa}, Fabian and {van Mulbregt}, Paul and {SciPy 1. 0 Contributors}},
        title = "{SciPy 1.0: fundamental algorithms for scientific computing in Python}",
      journal = {Nature Methods},
         year = 2020,
        month = feb,
       volume = {17},
        pages = {261-272},
          doi = {10.1038/s41592-019-0686-2},
archivePrefix = {arXiv},
       eprint = {1907.10121},
 primaryClass = {cs.MS},
       adsurl = {https://ui.adsabs.harvard.edu/abs/2020NatMe..17..261V}
}
%

\begin{appendix} 
\section{Additional figures}\label{sec:apx}
In this Appendix we provide two additional figures. Figure \ref{fig:HRD} shows the progenitor temperatures, luminosities and radii for the fiducial selection of BH progenitors where the BH mass exceeds 38\,M$_{\odot}$. Figure \ref{fig:core} shows the carbon-oxygen core versus BH masses for the selected models.

\begin{figure}
\centering
\includegraphics[width=0.99\columnwidth]{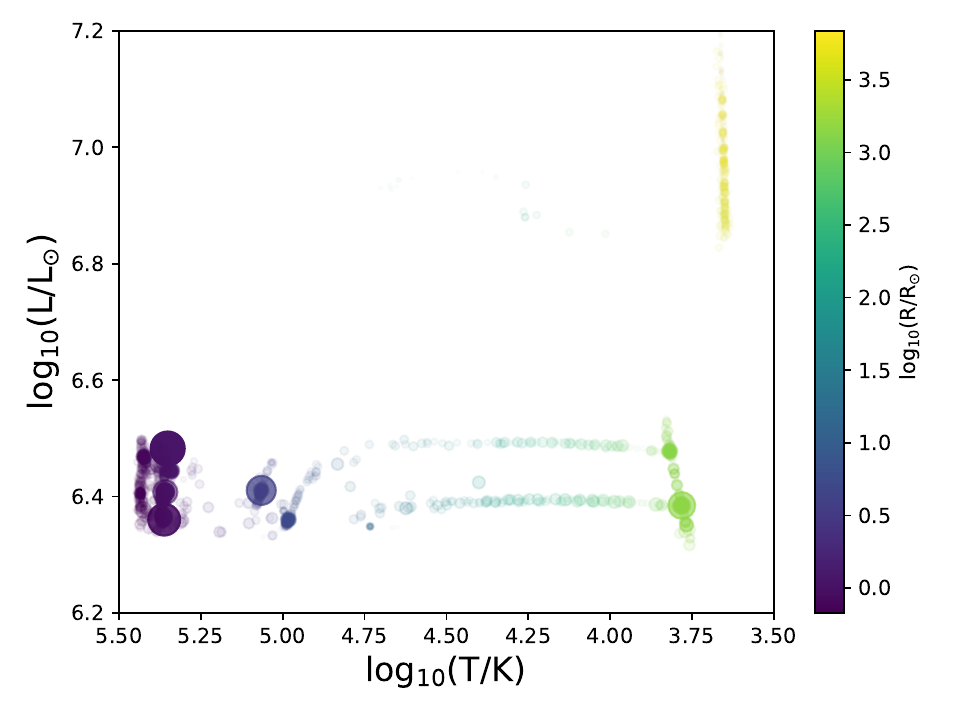}
\caption{Temperature-luminosity diagram for the fiducial selected models as described in Section \ref{sec:rates}. The size of the points corresponds to the model weighting, and the colour to the final radius of the star. Most of the progenitors are Wolf-Rayets or blue supergiants, with a smaller contribution from red supergiants (including those above the pair-instability mass gap). }
\label{fig:HRD}
\end{figure}

\begin{figure}
\centering
\includegraphics[width=0.99\columnwidth]{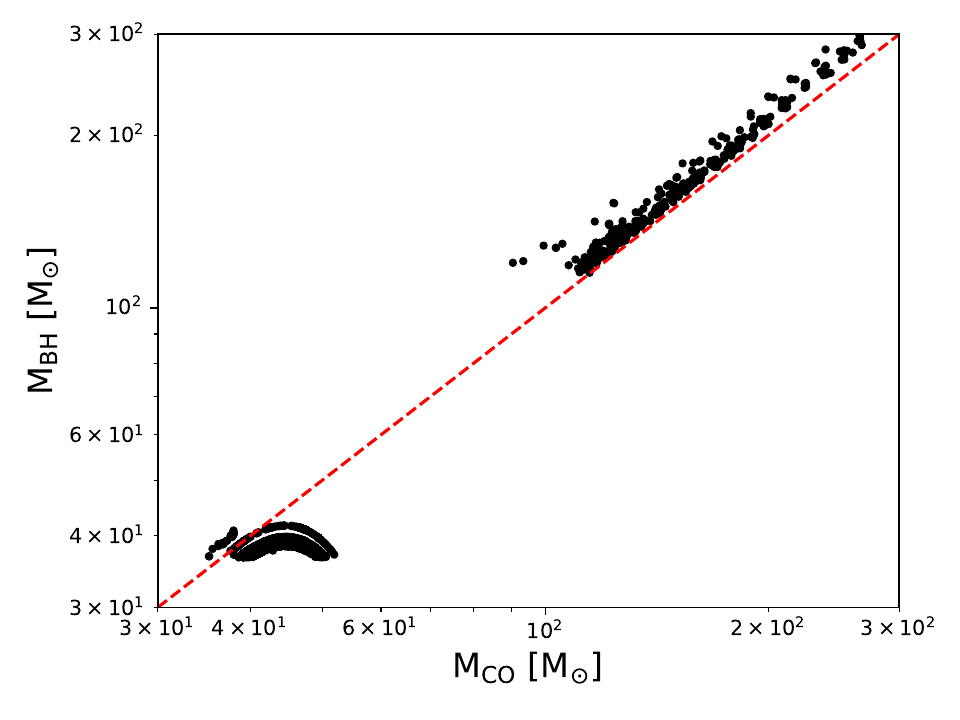}
\caption{Carbon-oxygen core masses versus the final remnant mass, for the fiducial selected models as described in Section \ref{sec:rates}. The red dashed 1:1 line is a reasonable approximation of the \citet{2012ApJ...749...91F} remnant mass prescriptions above $\sim$10\,M$_{\odot}$ \citep[see also][]{2023MNRAS.520.5724B}. The {\sc bpass} remnant masses deviate from this by no more than 10--20\% level below the pair instability mass gap, and track it well above.}
\label{fig:core}
\end{figure}

\section{Disc radii}\label{sec:apx2}
To calculate the initial disc radii - for input to the model of \citet{2020MNRAS.492.5655M}, \citet{2025MNRAS.541..429M} - we first define M$_\star$=M$_{\rm BH} +$M$_{\rm leftover}$. The initial disc radius can be estimated from angular momentum and mass conservation as the star is ``turned'' (from some process we do not specify) into a Keplerian flow. We therefore need an estimate of the {\it maximum} angular momentum available to the disc. On energetics grounds we can write for the pre-collapse star,
\begin{equation}
 -\frac{G M_{\star}^{2}}{ R_{\star}} + \frac{1}{2}  M_{\star}  v_{\rm crit}^{2} = 0
\end{equation}
where $v_{\rm crit}$ is the maximum pre-collapse stellar rotational velocity. The corresponding specific angular momentum of the star $j_{\rm crit}$ is approximately $\sim R_{\star} v_{\rm crit}$. Substituting into equation B.1 (and ignoring constants of order unity) we find
\begin{equation}
 -\frac{G M_{\star}}{ R_{\star}} + \frac{j_{\rm crit}^{2}}{R_{\star}^{2}} \approx 0
\end{equation}
and therefore
\begin{equation}
j_{\rm crit} \approx \sqrt{G M_{\star} R_{\star} } .
\end{equation}
The angular momentum in the accretion disc has to be less than or equal to the total pre-collapse angular momentum, therefore
\begin{equation}
M_{\rm leftover} j_{\rm disc} \leq M_{\star} j_{\rm crit}
\end{equation}
Where $j_{\rm disc} = \sqrt{G M_{\rm BH} r_{0}}$ and $r_{0}$ is the initial disc radius. We therefore find an absolute maximum\footnote{In principle the initial disc radius could be larger than this if the disc material somehow got all of the initial stellar angular momentum but not all of the leftover mass. We ignore this possibility here.} initial disc radius of 
\begin{equation}
r_{0, {\rm max}} \approx R_{\star}\bigg(\frac{M_{\star}}{M_{\rm leftover}}\bigg)^{2} \bigg(\frac{M_{\star}}{M_{\rm BH}}\bigg)
\end{equation}
in the extreme case that (i) the star is maximally rotating and (ii) the total pre-collapse angular momentum is transferred entirely to the disc. Since $j_{\rm crit} \propto v_{\rm crit}$ and  $r_{0} \propto j_{\rm crit}^{2}$ we can introduce factors to allow for sub-critical rotation velocities $v_{\star}$, and the fraction of angular momentum which goes into the disc $f_{\rm j,disc}$, as follows:
\begin{equation}
r_{0} \approx f_{\rm j,disc}^{2} R_{\star} \bigg(\frac{v_{\star}}{v_{\rm crit}}\bigg)^{2} \bigg(\frac{M_{\star}}{M_{\rm leftover}}\bigg)^{2} \bigg(\frac{M_{\star}}{M_{\rm BH}}\bigg)
\end{equation}
Finally, we adopt $0.1v_{\rm crit}$ for the pre-collapse rotational velocity and introduce the fraction of leftover mass which is accreted $f_{\rm acc}$, yielding
\begin{equation}
r_{0} \approx 0.01 f_{\rm j,disc}^{2} R_{\star} \bigg(\frac{M_{\star}}{f_{\rm acc} M_{\rm leftover}}\bigg)^{2} \bigg(\frac{M_{\star}}{M_{\rm BH}}\bigg)
\end{equation}
as adopted in Section \ref{sec:plateau}. 

The disc then must expand, as angular momentum is conserved while the mass of the flow drops (as some material is added to the black hole). To be precise, the quantity 
\begin{equation}
    M_{\rm disc}(t) \sqrt{R_{\rm disc}(t)} = {\rm constant} \approx f_{\rm acc} M_{\rm leftover} \sqrt{r_0}
\end{equation}
is conserved. The mass in the disc drops with time in a self similar fashion, as the accretion rate 
\begin{equation}
    \dot M_{\rm acc} \propto t^{-n}, 
\end{equation}
at large times \citep[e.g.][]{Pringle81}, such that 
\begin{equation}
    M_{\rm disc}(t) = M_0 - \int\dot M_{\rm acc}(t')\, {\rm d}t' \propto t^{1-n}, 
\end{equation}
and thus 
\begin{equation}
    R_{\rm out}(t) \propto t^{2n-2}.
\end{equation}
Of course, this expansion of the disc can be suppressed if angular momentum is lost from the disc -- i.e., if angular momentum is transported to large radii by winds. It is important therefore that the radii at large times (in this model) are larger than that inferred from observations. 
\end{appendix}


\end{document}